\documentstyle[12pt,aasms4,psfig]{article}

\def\,{\thinspace}

\newcommand{\solm}{M$_{\odot}$}

\newcommand{\rf}{\par\noindent\hangindent 15pt {}}

\newcommand{\vol}[1]{1}

\lefthead{Eckart et al.}
\righthead{Stellar Orbits Near Sgr~A*}

\begin{document}

\title{Stellar Orbits Near Sagittarius~A*}
\author{A. Eckart}
\affil{I.Physikalisches Institut, Universit\"at zu K\"oln,
Z\"ulpicher Str.77, 50937 K\"oln, Germany}
\and 
\author{R. Genzel, T. Ott, R. Sch\"odel}
\affil{Max-Planck-Institut f\"ur extraterrestrische Physik (MPE),
D-85740 Garching, Germany}
\received{}
\begin{abstract}
\noindent
The SHARP/NTT stellar proper motion data now cover an interval from 
1992 to 2000 and allow us to determine orbital accelerations 
for some of the most central stars. 
We confirm the stellar acceleration measurements 
obtained by Ghez et al. (2000) with NIRC at the Keck telescope.
Our analysis differs in 3 main points from that of Ghez et al.:
1) We combine the high precision
but shorter time scale NIRC/Keck data 
with the lower precision but longer time scale
SHARP/NTT data set;
2) We statistically correct the observed accelerations for geometrical
projection effects;
3) We exclude star S8 from the analysis of the amount and position of
the central mass.
\\
\\
From the combined SHARP/NTT and NIRC/Keck data sets we 
show that the stars S2, and most likely  S1 and S8 as well, 
are on bound, fairly inclined ($60^o<i<80^o$), and eccentric ($0.4<e<0.95$) 
orbits around a central dark mass.
The combination of both data sets results in 
a position of this central mass of 
48$^{+54}_{-24}$~$mas$ E and 
18$^{+42}_{-61}$~$mas$ S 
of the nominal radio position of Sgr~A*.
The mean statistically corrected enclosed mass 
derived from accelerations is $M_{acc}=(5 \pm 3) \times 10^6$\solm
with current radial separations of S1 and S2 
from SgrA* of about $8-10~mpc$.
This enclosed mass estimate is derived from
{\it individual} stellar orbits as close to the massive 
black hole at the center of the Milky Way as currently possible.
Although the uncertainties are large this estimate 
is fully consistent with the enclosed 
mass range of $(2.6-3.3)\times10^6$\solm~derived by 
Genzel et al. (2000) from radial and/or proper motion velocities
of a homogenized sample of sources. 
\\
\\
Star S8 was excluded from the analysis, since for the current proper 
motion velocity and radial separation from the center we find that the 
measured acceleration  
requires orbital motion around a compact object with a mass 
in excess of $3\times10^6$\solm.
The data suggest that this star 
{\it either} was or is subject to a close interaction with a 
different object
{\it or} that its position measurements are influenced by the emission of a 
different cluster star. 
Therefore, we base the analysis of the 
enclosed mass solely on the available data for S1 and S2.
We also discuss two late type stars with 
projected separations from SgrA* of about 0.5'' and 1''.
In addition to proper motions these stars have known 
radial velocities. Orbit calculations indicate that
those stars are very likely at larger physical distances from 
the center and part of the larger scale central stellar 
cluster with a core radius of approximately $0.3~pc$.

\end{abstract}
 
 
\section{
\label{sec1}
INTRODUCTION}
\normalsize 
High resolution near-infrared imaging and spectroscopy with large 
telescopes resulted in a determination of the amount and 
concentration of mass at the center of the Milky Way 
(Sellgren et al. 1990, 
Krabbe et al., 1995, 
Haller et al. 1996, 
Eckart \& Genzel 1996, 
Genzel et al. 1998, 
Ghez et al. 1998, 
Genzel et al. 2000).
Diffraction limited images, proper motions and most recently
the detection of acceleration of stars in the vicinity of the 
compact radio source Sgr~A$^*$ (Ghez et al. 2000, Eckart et al. 2000a)
have led to the conclusion that 
this source is associated with a central black hole with a mass of
about 3$\times$10$^{{ 6}}$ \solm.
Using the MPE speckle camera SHARP at
the 3.5 m New Technology Telescope (NTT) of the European
Southern Observatory (ESO) from 1992 to 2000
we have been conducting a program to study the properties
of the central nuclear stellar cluster via near-infrared high
spatial resolution measurements.
This program has resulted in the very first detection of proper 
motions of stars that correspond to velocities of up to 1400~km/s 
in the central arcsecond in the vicinity of Sgr~A$^*$ 
(Eckart \& Genzel 1996, 1997). 
These results had 
been confirmed by Ghez et al. (1998) using NIRC on the Keck telescope.
On the 1.6$\sigma$ to 3$\sigma$ level 
we have now detected orbital curvatures of the stars S1, S2, and S8
which confirm the recent results by Ghez et al. (2000).
\\
\\
Over the past 8 years 
the observed motions translate into a two dimensional velocity dispersion 
of the stars in the central arcsecond (corrected for the measurement error) 
of the order of 600 km/s.
Overall the stellar motions do not deviate strongly from
isotropy and are consistent with a spherical isothermal stellar cluster
(Genzel et al. 2000).
However, a small deviation from isotropy is found for the sky-projected 
velocity components of the young, early type stars. 
Most of the bright HeI emission line stars are on tangential orbits. 
This overall rotation could be a remnant of the original angular 
momentum pattern in the interstellar cloud from which these stars were formed.
The projected radial and tangential proper motions of the
fainter, fast moving stars within $\approx${}1'' from SgrA* (the 
`SgrA* cluster') suggest that they may largely be moving on 
 very eccentric  orbits. 
Speckle spectroscopy with SHARP at the NTT 
(Genzel et al. 1997) and slit spectroscopy with 
ISAAC at the VLT (Eckart, Ott, Genzel 1999, Figer et al. 2000) 
suggest that several of them are early type stars. 
This is consistent with the idea that these stars are members of the early 
type cluster with small angular momentum and therefore fell into the 
immediate vicinity of SgrA* (Genzel et al. 2000, Gerhard 2000,
see also discussion in Ghez et al. 2000).

The detection of orbital curvature for 3 central stars now allows us
for the first time to constrain their orbital elements and to investigate 
the central mass distribution at the smallest currently accessible separations
from Sgr~A$^*$.
In the following subsection \ref{sec1.1} we give a summary of why it is 
worthwhile to determine stellar orbits near the dark mass at the Galactic Center
and give a general description of how that can be done best. 
In section \ref{sec2} we present the new proper motion and acceleration data
and derive the position of the central dark mass.
In section \ref{sec3} we show how the acceleration measurements can be used
to obtain an enclosed mass estimate.
In section \ref{sec4} we then make use of the enclosed mass and its position 
and discuss possible orbital solutions for several of the central stars.
Finally a summary and conclusions are given in section \ref{sec5}.

\subsection{
\label{sec1.1}
Stellar orbits and the central mass distribution} 
\normalsize 

The point source sensitivities currently reached in near infrared observations
of the central stellar cluster are of the order of 
$K\sim16-17$ at an angular resolution of 50 - 150 mas
(Eckart \& Genzel 1996, Ghez et al. 1998, Genzel et al. 2000, Ghez et al. 2000).
These current limits are mainly a result of the imaging techniques at the 
diffraction limit and not due to the sensitivity of the instrumentation.
With these values the analysis of the central mass distribution
for separations of less than 0.5'' ($\sim$20~mpc) from the radio point 
source SgrA* has to be based on detailed measurements of individual 
stellar orbits. We assume that the trajectories of stars can be approximated 
by Keplerian shaped orbits around a central mass.

Relative Keplerian orbits are described by 6 orbital elements: 
three angles, the eccentricity $e$, the semi-major axis
$a$ and a reference time $T$ for a defined position (i.e. the periastron).
Currently for a few of the early type high velocity S-sources as well as
two late type stars that are close to the center in projection, our 
detailed orbital analysis (section \ref{sec4}) can only be based on 5 known 
quantities that constrain the orbits. 
In addition to the projected positions and proper motions these are 
the orbital curvatures (for the S-sources) and the line of sight 
velocities (for the late type stars).
However, for visual binaries - as is the case for a combination of the
mass at the center of the Galaxy and any of the fast moving stars 
in its vicinity - all these orbital elements can be derived from a sufficiently 
large section of the orbit, except for the sign of the inclination. 
This has to be inferred from Doppler shifts via spectroscopy. 
The visual binaries situation holds if the exact position
of the black hole on the near-infrared images is known
(see Menten et al. 1996). 
The orbital period and central mass can then be determined
via Kepler's third law and the equation of motion i.e. from Newton's law.
Here we present a first step to reach that goal using the combined 
SHARP/NTT and NIRC/Keck data sets.

The orbital analysis in the presence of a possibly partially 
extended central mass has to await the detection of fainter stars 
($K\ge 17$) closer ($<2~mpc$) to SgrA* or the availability of long 
period measurements (of the order of 100 years or more; see section 
\ref{sec4.2}) of the presently known high velocity stars within 
the central arcsecond.  
The expected orbits have been discussed in the context of the
Galactic Center by Rubilar \& Eckart (2001, see also
Fragile \& Mathews, 2000, Munyaneza, Tsiklauri \& Viollier, 1998, 
and Jaroszy\'nski, 1998).
The shortest orbital time scales that could be observed in the
near future will be of the order of only a few months.
These measurements are well within reach 
using the new instrumentation (VLT, Keck and LBT interferometers; see 
Rubilar \& Eckart 2001, Eckart, Ott, Genzel 2001a, 2001b).

\section{
\label{sec2}
ACCELERATIONS} 
\normalsize

From stellar orbital accelerations we can determine the position of the 
central mass. They can also be used to derive an additional estimate 
of the enclosed mass, help to constrain the stellar orbital parameters,
and to show whether or not the high velocity stars in the vicinity of SgrA* 
are bound.
In the following we first present the results from the SHARP/NTT data and 
then describe how we determine the position of the dark mass from the combined
SHARP/NTT and NIRC/Keck data sets.

\subsection{
\label{sec2.1}
Stellar accelerations and proper motions} 
\normalsize 
In order to derive accelerations a precise measurement of the change in 
proper motion velocities is required. Such a measurement can only be 
performed for stars that are bright, well separable from each other, and
close enough to the center in order to show detectable accelerations. 
At the Galactic center the stars S1, S2, and S8 fulfill this 
requirement and a curvature of their trajectories has already been detected by
Ghez et al. (2000). 
In order to estimate stellar accelerations from the SHARP/NTT data
we have used three different methods:
a) variation of proper motion velocities as a function of time,
b) parabolic fits to the time position plots and
c) fitting Keplerian orbits to the observed data.
The results of all three methods are in good agreement and give for stars
S1, S2, and S8 accelerations similar to those reported by Ghez et al. (2000).
In the following we discuss properties and results of the three methods 
in more detail.
\\
\\
We prefer to estimate the stellar accelerations via the 
{\it variation of proper motion velocities} since this method
stays closest to the data. 
The only requirement is that over a period of 3 to 4
years a linear fit to the data results in reliable estimates of proper motions.
The SHARP/NTT proper motion data set covers the time interval from 1992 
till 2000 and allows us to derive and confirm  for the stars 
S1, S2, and S8 curvatures
of their orbits on the 1.6$\sigma$ to 3$\sigma$ level.
In Fig.~\ref{fig01} (see also Tab.~\ref{t01})
we show the offset positions from SgrA* 
in declination and right ascension as a function of time.
In the case of S1, S2, and S8 we compare them to parabolic functions
(see below).
Linear fits to the first and second half of the
data sets for these three stars are shown by
Eckart, Ott, Genzel (2001a, 2001b) and clearly
result in different slopes over successive time intervals.
This difference in slopes
divided by the time difference between the two intervals is
a direct measure of the orbital curvature. 
We divided the data into an early (1992 - 1996) and a late epoch (1997 - 2000).
For each epoch we calculated three velocities from data in different
sub-intervals. 
Individual accelerations were then estimated from quotients of
all combinations of velocity and time differences between the early and 
late epoch. We verified the method by comparing the results to those obtained
by the other two methods. We also applied it to  the data shown 
in Fig.2 in Ghez et al. (2000) and reproduced the acceleration values
quoted in their Table.1 for the time averaged epoch
\footnote{ \footnotesize{ Our assessment is that the acceleration 
values quoted in Table.1 by Ghez et al. (2000) 
must correspond to the time averaged epoch 1997.6 rather the
epoch 1995.53 as mentioned in the corresponding figure caption.
Combining the high proper motions of S1 and S2 
with the 1995.53 positions - these accelerations would otherwise result in a 
position of the central mass approximately 0.15'' further north 
compared to what is shown in their Fig.3}}
 1997.6.
In Tab.~\ref{t01} we compare the total accelerations and their 1$\sigma$
uncertainties for the three stars S1, S2, and S8 as derived from 
the SHARP/NTT and the NIRC/Keck data.
The data agree to within less than 3.3$\sigma$.

A {\it parabolic fit} 
is adequate as long as the acceleration is nearly
constant over the observed orbit segment. This approximation is not
always justified for the entire 8 years of SHARP/NTT observations
analyzed here.
In particular, star S2 may have completed about half of its orbit 
during this period.
Depending on the observed orbital 
section a parabolic fit may therefore result in
an inaccurate estimate of the acceleration.
We have, however, used this method for the purpose of comparison.
In Fig.\ref{fig01}
we show the data together with parabolic functions that have curvatures
as listed in Tab.\ref{t01}.
For stars S1, S2 parabolic fits give 
similar results. For S8 parabolic fits indicate a curvature closer to 
$2.3 mas~yr^{-2}$ (as plotted in Fig.\ref{fig01})
rather than $3.3 mas~yr^{-2}$ which is, however, well 
within the errors of the value derived above.

{\it Fitting Keplerian orbits to the data} has the advantage that 
it provides the appropriate dependence of the projected positions 
as a function of time.
It has the disadvantage that it assumes the presence of Keplerian orbits
and therefore requires knowledge of the enclosed mass and mass density.
In section \ref{sec4} we show that the enclosed mass and mass density estimates
derived from accelerations are in agreement with a compact 
$3\times10^{6}$\solm ~central object.
In section \ref{sec4} we make use of this fact and fit Keplerian orbits
to the data.
For the stars S1 and S2 the range of Keplerian accelerations is in close 
agreement with our estimates from the other two methods. 
For S8 the observed acceleration
lies well above the prediction from Keplerian orbits. Reasons for that are
discussed in detail in section \ref{sec4.2.2}.

{\it Determination of Proper motions:}
In the presence of significant orbital curvature a precise knowledge 
of positions and proper motion velocities for a given reference epoch 
is required in order to carry out orbit calculations.
As reference epochs we chose the centers of the time intervals 
covered by the SHARP/NTT and NIRC/Keck data sets.
In Tab.~\ref{t02} we list the positions that we derived for those 
epochs from our own data and the data presented by Ghez et al. (2000). 
In both cases the near-infrared positions are measured relative to the 
radio position of SgrA* (Menten et al. 1996).
For S1 and S2 the accelerations given in Tab.~\ref{t01} correspond to a 
yearly change of the proper motion velocity of about 80-160~km/s.
This is of the order of both the mean measurement uncertainty and the
mean differences in proper motion velocities between 
the SHARP/NTT and NIRC/Keck data sets (with a time difference of about 
1 year between central epochs;
Eckart \& Genzel 1996, 
Ghez et al. 1998, 
Genzel et al. 2000, 
Ghez et al. 2000).
In order to combine the SHARP/NTT and NIRC/Keck data sets
and to obtain higher precision proper motion values for orbital
calculations we have used our acceleration values and the proper 
motion velocities for epoch 1995.4 by Ghez et al. (1998) to derive 
mean proper motion 
velocities for the two reference epochs 1996.5 for the SHARP/NTT data set 
and 1997.6 for the NIRC/Keck data set 
(listed in Tab.~\ref{t03}; see also tables in the Appendix).

We have also investigated the proper motion data of other stars 
within the SgrA* cluster. The number of further candidates that would allow 
us to detect curvature from the present SHARP/NTT data set is small.
Star S3 apparently became fainter and disappeared - stars S4, S5, and S6 
are in a crowded area with S4 even being multiple - and S9 is too faint 
and is approaching S10 (see notes in appendix of Genzel et al. 1997).
In addition to S1, S2, and S8 the best candidates for detecting 
orbital curvature, in terms of their
brightness and source crowding, are S7, S10, S11.
In Fig.~\ref{fig01} we show the proper position time diagrams for 
these three stars and list their separations from Sgr~A$^*$ and
upper limits for their accelerations in Tab.~\ref{t01}. 
 
In addition to the central  early type stars there are two late type stars 
with projected separations of only about 0.5'' and 1'' listed in Tab.\ref{t01}
as S18 and star No.25, respectively.
Those stars are discussed in detail in section~\ref{sec4}. 
In  Fig.~\ref{fig02} we show the 
proper position time diagrams for these two stars that show no significant 
curvature.

\subsection{
\label{sec2.2}
The Position of the central mass} 
\normalsize 
The accelerations are consistent with recent results by Ghez et al.(2000)
and imply that the three stars orbit a central, compact mass.
For each of the two sources S1 and S2 the
acceleration values define an acceleration vector at an angle $\phi$
that should point towards the central source. 
Here we assume that the probability for the location of the
central mass is uniform in $\phi$.
In Fig.~\ref{fig03} the stars 
S1 and S2 have been plotted at their time 
averaged position resulting from the corresponding data sets.
The measurement uncertainties define an error cone. 
For the presentation in Fig.~\ref{fig03} the 
dashed and dotted lines indicate an 
error cone that corresponds to a width of 2$\sigma$ in 
deviation from the nominal 
direction indicated by the acceleration vector.
In order to determine the location of the central mass 
we perform a maximum likelihood (ML) analysis.
As a ML score we use 

\begin{equation}
\label{eq00}
log(ML) =  - \chi_{S1}^2/2 - \chi_{S2}^2/2~~.
\end{equation}

The thin contour lines in Fig.~\ref{fig03} indicate the 
locations at which $log(ML)$ 
drops by 0.5 below the corresponding peak values.
Here $\chi^2 = (\phi-\phi_0)^2/(\Delta \phi)^2$, $\phi_0$ denotes the 
angle of the acceleration vector, $\phi$ the angle
of any radial line within an error cone, and
$\Delta \phi=\sigma$ the half width of the cone.
The central filled circle in Fig.~\ref{fig03} marks the radio position 
of SgrA* and corresponding uncertainties of $\pm$30~mas.
Using the observed curvature value and the enclosed mass range 
of 2.6 to 3.3$\times$10$^6$\solm ~imposes a limit on the 
projected distance of S1 and S2 from SgrA*. 
For the SHARP/NTT data this leads to an improvement in the 
determination of the SgrA* position. We account for this effect in 
Fig.~\ref{fig03} by multiplying the $log(ML)$ scores 
of the error cones with a Gaussian prior of the appropriate 
$1/e$ width centered on the time averaged positions of S1 and S2. 
From the projection of the 1~$\sigma$ contour line 
(thin contour line E of the center in Fig.~\ref{fig03})
the multiplied probabilities derived from the
SHARP/NTT and NIRC/Keck 
data result in a position of a central dark mass of 
48$^{+54}_{-24}$~$mas$ E and 
18$^{+42}_{-61}$~$mas$ S
of the nominal radio position of Sgr~A*.
Within these limits the central mass is located at
the 68.5\% confidence level ($\Delta \chi = 1.0$).
At the 90\% confidence level
($\Delta \chi = 2.71$)
the central mass is located 
in an interval given by
48$^{+109}_{-48}$~$mas$ E and 
18$^{+72}_{-133}$~$mas$ S (thick contour line E of the
center in Fig.~\ref{fig03}).  
Fig.~\ref{fig03} shows that at the current  $\pm$30~$mas$  
uncertainty of the radio position of Sgr~A*
the presently available accelerations of stars S1 and S2 alone
are fully consistent with the hypothesis that the radio 
source SgrA* is coincident with the center of the dark mass.

\section{
\label{sec3}
THE ENCLOSED MASS} 
\normalsize 
Assuming Keplerian orbits the 
observed projected stellar accelerations allow us only to derive lower
limits 
to the enclosed mass. By inferring an estimate of the true physical 
separation of the stars from the center these lower limits can be 
statistically corrected for projection effects and compared to previous 
estimates of the amount and compactness of the enclosed dark mass.
In section \ref{sec3.1} we first determine the volume that contains 
the three stars S1, S2, and S8, describe in section \ref{sec3.2.1} how we use
this information to correct for projection effects,
verify the validity of that method in section \ref{sec3.2.2}, and 
then apply it in section \ref{sec3.2.3} to the observed data.

\subsection{
\label{sec3.1}
Location of stars relative to Sgr~A$^*$} 
\normalsize 
In Fig.~\ref{fig04} we plot the quantities that are relevant in 
describing the position of stars relative to the center of the
Galaxy. 
The line of sight position of individual stars with respect to
the plane of the sky containing Sgr~A$^*$ is unknown.
A correction for a geometrical projection of the acceleration values 
can, however, be carried out 
statistically by simply estimating the volume that 
contains the central stars S1, S2, and S8.
In order to derive an estimate of this volume we may use the 
proper motion velocities and projected positions of these stars 
in comparison to the velocity dispersions and number density as a 
function of the true (non projected; 3-dimensional) radius $r$.
The central fast moving stars have significantly curved orbits.
It is therefore unlikely that
these fast moving stars are objects on highly elliptical
orbits ($e>1$) that just happen to fly by the central position on 
almost linear trajectories.
In the following we argue 
that the sources S1, S2, and S8 are likely located within a 
spherical volume of radius 15$\pm$3~$mpc$ (0.4''$\pm$0.1'').
 
{\it Estimate from velocities and positions:}
In the following $R$ denotes the projection of the 3-dimensional
separation $r$ of the star from the center
and $V_{star}$ the projection of the  3-dimensional
velocity $v_{star}$ of the star.
An estimate of the size can be obtained 
by comparing the proper motion velocities $V_{star}$ 
to the 3-dimensional velocity dispersion 
obtained from  the parameterized functional form
of the true number density distribution $n(r)$ 
and the true radial and tangential velocity dispersions 
$\sigma_r(r)$ and $\sigma_t(r)$ of the best anisotropic 
Jeans model (Genzel et al. 2000).
Those quantities are integrated along the line of 
sight at the projected separation $R$ from Sgr~A* of 
the individual stars (formulae 5 and 6 in Genzel et al. 2000).
This way we use the projected position and velocity information
for each of the stars simultaneously. 
Both  $V_{star}$ and  $R_{star}$ are lower limits to 
$v_{star}$  and $r_{star}$.
For the probability of a star to exhibit a proper motion velocity $V$ 
in excess of $V_{star}$ we can therefore write
\begin{equation}
\label{eq01a}
P(V>V_{star},R) \ge P(V>V_{star},r) \ge P(V>v_{star},r)~~.
\end{equation}

For a given velocity dispersion
 $\sigma^2(r) = \sigma_r^2(r) + 2 \sigma_t^2(r)$ 
the probability $P(V>V_{star},r)$ can be calculated via

\begin{equation}
\label{eq01b}
P(V>V_{star},r) = 1- P(V \le V_{star},r) = 
   1 - 1/\sigma^2(r)\int_0^{V_{star}} v\exp(-v^2/(2\sigma^2(r))) dv~~.
\end{equation}

The velocity dispersion $\sigma(r)$ and therefore also the
probability $P(V>V_{star},r)$ decrease with increasing radii $r$.
For fast stars it becomes increasingly unlikely that they belong to 
statistical samples at correspondingly larger radii.
Therefore, we interprete $P(V>V_{star},r)$ as a measure of how likely 
it is that the star belongs to a sample of stars 
at that radius $r$ or larger.
Using $R$ instead of $r$ we can calculate 
$P(V>V_{star},R)$ as an upper limit of this probability
$P(V>V_{star},r)$.
The mean probability of the three stars S1, S2, and S8 to belong to 
samples of stars at the corresponding radii 
$R_{S1}$, $R_{S2}$, and $R_{S8}$
- or larger - is only about $P_{init}=$33\%.
This implies that the mean probability of these three sources 
to belong to samples of stars at their true 3-dimensional separations  
$r_{S1}$, $r_{S2}$, and $r_{S8}$ - or larger - is even less than that.
The value of $P_{init}$ drops by a factor of 2 (i.e. to the FWHM 
value of that probability) at a mean radius of $r=13.7~mpc$ 
and by a factor of 
three at a radius of $r=14.8~mpc$. 
The probability of the stars to belong to samples at even larger radii
lies well below 10\%.
We therefore adopted a value of about 15~$mpc$ (0.4``).
as a reasonable estimate of the radius of the volume that contains 
all three stars.

{\it A safe lower bound} to the size estimate of the volume 
described above is given by the
upper limit of the projected separation of the stars from SgrA$^*$.
Of the three stars S8 has the largest projected separation from the 
center (see Tab.~\ref{t02}).
Therefore we adopt 12~$mpc$ (0.3'') as a lower bound to the radius of the
volume containing the three stars. This limit compares 
favorably with the size estimate derived above.

\subsection{
\label{sec3.2}
Enclosed mass estimates from accelerations}
\normalsize 
 
\subsubsection{
\label{sec3.2.1}
Correction of accelerations for projection effects}
\normalsize 
 
For a star at a projected separation $R$ from 
the center and a total enclosed mass M 
one can calculate
the projected, observable acceleration $a_{obs}$ via
\begin{equation}
\label{eq02}
a_{obs} = G M cos^3(\theta) R^{-2}~~~.
\end{equation}
Here $\theta$ is the angle between the radius vector to the star and the 
plane of the sky containing the central mass.
Plotting the lower limits $M cos^3(\theta)$ of the enclosed mass 
as a function of the 
projected radius is equivalent to the assumption that the stars 
are in exactly the same plane of the sky as the central dark mass 
at the position of Sgr~A$^*$. 
This assumption is not justified and the 
 approach does not answer the question of whether or not the 
observed projected accelerations are in agreement with the value 
and compactness of the enclosed mass derived at larger radii with 
different methods (Genzel et al. 2000, Ghez et al. 1998). 
A more realistic approach needs to correct for  
geometrical projection effects. 
A statistical estimate of $M$ can be derived by using median values.
As a consistent error estimate we use the median error defined as the 
median of the deviations of the individual estimates from their median.
The quantity $(cos \theta)^{-1}$ increases monotonically with the distance 
from the plane of the sky in which Sgr~A$^*$ is located and its median
can be calculated under the assumption of a stellar density 
distribution $n(r)$.

Using median values and a volume derived for a ensemble of
stars make this method of correcting for geometrical effects
much less susceptible to extreme correction values that occur
for instance in the case of stars with large physical separations and
small projected separations from SgrA* (i.e. $\theta$ approaching $\pi$/2).
Contributions from those values would become dominant in case of
a calculation of an expectation value for $(cos \theta)^{-1}$ using
$n(r)$ values as a weights.
 
\subsubsection{
\label{sec3.2.2}
Validity of the approach}
\normalsize 
 
In order to verify that the above described method results in acceptable
statistically corrected enclosed mass estimates we performed simulations.
The results of the de-projection procedure
 - presented in the $R-log(M)$-plane -
show that the distribution of the de-projected mass estimate tends 
to under-estimate $M$ for flat stellar distributions, whether the 
central mass is assumed to be point-like or extended, but is tightly peaked 
around the true value of $M$ for steep cusp-like distributions.

We assumed a sphere with radius $r=15~mpc$ and a stellar 
number density $n(r)$ surrounding the dark mass of 
3$\times$10$^6$\solm. 
For each star (we used several 1000) at a separation $r$ from 
the center and a total enclosed mass M we calculated the projected radius 
R and the projected, observable acceleration $a_{obs}$ via
equation~(\ref{eq02}).
In Fig.~\ref{fig05}a,b,c we show the density of 
data points in the R-logM-plane for combinations of
a central point mass (BH) or an
extended central mass distribution 
and a constant or cusp-like
stellar number density distribution.
Almost all data points underestimate the enclosed mass 
$M$ and the value of that
estimate drops dramatically towards smaller values of R.
The density increase of data points towards larger projected distances from
the center is due to the fact that the projection effects decrease for 
stars towards the projected edge of the limited volume. 
In Fig.~\ref{fig05}d,e,f we correct both the projected 
radius R and the upper limit of 
the mass using the formalism outlined above in section \ref{sec3.2.1}. 
A higher density of points is located along the correct value of the enclosed 
mass and the remaining estimates are almost equally distributed 
above and below that value.

The available number density counts provide some evidence for 
an increased volume density of stars towards the center 
(Genzel et al. 2000, Alexander 1999, see also Alexander \& Sternberg 1999).
In Fig.~\ref{fig05}b and Fig.~\ref{fig05}e we have calculated the 
expected projected and statistically corrected mass estimates using 
a $r^{-7/4}$ stellar density law as an extreme case. 
The corrected mass estimates spread almost symmetrically 
about the expected value.
At any projected radius the stellar number density along the line of sight 
is now biased towards the plane of the sky that contains SgrA*.
This results in enclosed mass estimates that are less affected by 
the geometrical projection.
For about 70\% of all stars in Fig.~\ref{fig05}b the projected 
enclosed mass estimate accounts for more than 70\% of the true mass value. 
This demonstrates that for steep cusps the majority of the projected 
mass estimates will be much closer to the true value than in the case 
of a constant density distribution. 
For a $r^{-7/4}$ density law we would expect at least for two stars 
a mass estimate of  at least $\sim$70\% of the enclosed mass.
This is only barely fulfilled by S8 and S2. This shows, that a 
larger number of stars with significant curvature will greatly 
improve our knowledge on the presence and nature of a central stellar cusp.

\subsubsection{
\label{sec3.2.3}
Application to the measured data} 
\normalsize 
In Fig.~\ref{fig06}a we show projected mass estimates derived from the 
observed accelerations as a function of the projected radius listed in 
Tab.~\ref{t01}. 
In addition we indicate the mass distribution obtained 
from stellar and gas dynamics (for $R=8.0$ kpc; see caption of
Fig.~\ref{fig06} and Genzel et al. 2000, 1997, 1996, 
Eckart \& Genzel 1996, 1997, Ghez et al. 2000,1998).

As expected from the simulations presented in Fig.~\ref{fig05} 
the estimates obtained from S1 and S2 fall well below the 
enclosed mass estimate of 3$\times$10$^6$\solm ~derived 
previously (see references above).
For S8, however, (see discussion in section ~\ref{sec4.2.2}) 
and especially for S7, S10, S11
(not shown in Fig.~\ref{fig06}a) - for which only {\it upper} 
limits of the acceleration could be obtained - the mass estimates are
well above  3$\times$10$^6$\solm. 
For S7, S10, and S11 these {\it upper} limits range between 
1.1$\times$$10^7$\solm and 1.6$\times$$10^7$\solm.
In Fig.~\ref{fig06}b we show the mass estimates 
for S1, S2, and S8 as derived from the observed 
accelerations and corrected for projection effects following the method
outlined in sections ~\ref{sec3.2.1} and ~\ref{sec3.2.2}.
The correction factors obtained for different volume sizes are listed in
Tab.~\ref{t04}. The derived corrected radii and mass estimates are given
in Tab.~\ref{t05}. 
The values cover a mass range of 2.9 to 7.2$\times$10$^6$\solm~ over
separations from Sgr~A$^*$ between 8 and 15~$mpc$.

We compare the data to enclosed mass estimates 
as a function of separation from Sgr~A$^*$ 
obtained assuming (physically not realistic)
Plummer like density distributions 
(see discussion in Genzel et al. 1996, 2000)
with a core radius  $r_c$  and a mass density $\rho(0)$ 
at the very center of the distribution.
We chose the exponent $\alpha$=5, since this corresponds to 
the steepest currently observed drop in cluster mass density.
For the stars S1 and S2 which are currently closest in projection to
SgrA$^*$ the mean value and error of the enclosed mass corrected for
a volume radius of about $15~mpc$ is 
$M_{acc}=(5 \pm 3) \times 10^6$\solm.
This value is fully consistent with an enclosed mass distribution that 
is flat down to radii of about 8~$mpc$ with a value of 
$3 \times 10^6$\solm~and a lower limit to the 
mass density of $3.7 \times 10^{12}$\solm$~pc^{-3}$ 
for a core radius of $r_c=5.8~mpc$ as previously derived from the proper 
motion data (Genzel et al. 2000).

As is apparent from Fig.~\ref{fig06}b this is the smallest range of true 
(not projected) separations from SgrA* for which a mass estimate 
corrected for projection effects has been derived so far.
The fact that $M_{acc}$ lies systematically above the enclosed mass obtained 
at larger radii can very likely be attributed to the fact that
the volume size has been estimated correctly but the stars 
are systematically 
closer to Sgr~A* along the line of sight than the median distance
at the given projected radius (see Fig.~\ref{fig04}).
Alternatively, the estimate of the volume radius in which 
the two stars S1 and S2 
are located is too large - which will result in the same 
effect and therefore in a correction 
that systematically over estimates the enclosed mass.
Orbit calculations assuming a $3\times10^6$\solm~point mass 
yield separations from the SgrA* plane of the sky of $6-7~mpc$ for S1 and S2.
This indicates that for a volume radius of $15~mpc$  
the described effect is in fact relevant.

Assuming a compact enclosed mass of $3 \times 10^6$\solm~the range of 
derived mass estimates can also be used to qualitatively judge its 
compactness. 
For comparison we plotted in Fig.~\ref{fig06}b 
the calculated Plummer like enclosed mass distributions for central
mass density $\rho(0)$ and core radius values $r_c$ 
of  $10^{13}$\solm$pc^{-3}$ and $4.2~mpc$ 
and $10^{14}$\solm$pc^{-3}$ and $1.9~mpc$, respectively. 
This comparison demonstrates that
\\
a) the estimates of the central enclosed mass and compactness derived from 
   acceleration measurements for stars S1 and S2 are fully consistent 
   with previously determined values (Genzel et al. 2000) and that
\\
b) under the assumption of a compact $3 \times 10^6$\solm ~central 
   dark mass the current acceleration data allow central 
   mass densities of $>10^{13}$\solm$pc^{-3}$ and core radii of $<4~mpc$.
   The star S2 currently (2000) is at a projected distance of 
   only about $60~mas$ from the center. 
   This is 4 times smaller than the minimum
   radius reached by the Jeans modeling (Genzel et al. 2000).
   If the orbit of S2 remains consistent with a
   compact mass of 3.0$\times$10$^6$\solm ~~the mass density 
   is at least 64 times higher than the value based on the Jeans modeling
   i.e. 2.4$\times$10$^{14}$\solm~pc$^{-3}$.
   In this case the collapse life time of a hypothetical cluster of dark
   mass would shrink to only a few 10$^6$ years (Maoz 1998).

\section{
\label{sec4}
STELLAR ORBITS CLOSE TO Sgr~A$^*$} 
\normalsize 

In the following we discuss possible Keplerian orbits for 
the three early type S-sources
S1, S2, and S8 as well as two late type stars: S18 and star No.25 in
Tab.1 by Genzel et al. (2000).
In the following section \ref{sec4.1} 
we first describe the algorithm we use to constrain the stellar orbits.
In sections \ref{sec4.2} and \ref{sec4.3} we then apply it to the combined 
SHARP/NTT and NIRC/Keck data sets of the three high velocity stars S1, S2, and
S8 and the two late type stars, respectively.

\subsection{
\label{sec4.1}
Orbit calculations}
\normalsize 

A complete global fit has to include the measurement 
errors of the relative positions and velocities
given in Tab.\ref{t02} and Tab.\ref{t03}, the uncertainties 
in the position of the central mass as well as its amount.
In order to get a first insight into the stellar orbits we first
restrict ourself to the case of a compact mass of
 3$\times$10$^6$\solm~and and a location of it that coincides with
the nominal position of SgrA*.
The influences of the uncertainties of these quantities on the 
3-dimensional orbits will be discussed in section \ref{sec4.3}.
 
We have chosen to present the results of our simulations in the 
$v_{z}$-$s_{z}$-plane rather than the semi-major axes and eccentricity
plane since 
this representation is closer to the observations.
Progress in diffraction limited near-infrared 
spectroscopy now allows ongoing experiments to determine the line of 
sight velocity of the central stars.
Calculated semi-major axes and eccentricities of the resulting
orbits are listed in Tab.\ref{t06}.
For two late type stars at the projected separations of about 0.5'' and 1''
from SgrA* radial and proper motion velocities are known. 
For these stars only the positions 
along the line of sight are undetermined.
For the stars S1, S2, and S8 the line of
sight velocities $v_{z}$ and positions $s_{z}$ 
are currently unknown. 
We considered orbits for the ranges 
of $-3500~<~v_{z}~<~3500~km~s^{-1}$ and $0~<~|s_{z}|~<~40~mpc$.
These intervals correspond to more than 5 times the central 
velocity dispersion and about twice the radius of the Sgr~A$^*$ cluster
and include all possible bound orbits.
To judge the quality of the orbital fits we calculated 
reduced $\chi^2$ values via

\begin{equation}
\label{eq04}
\chi^2=
\frac{1}{m-n}
\sum{\frac{(|{\bf x(t_i)-c(t_i)}|)^2}{\sigma^2}}~~.
\end{equation}
 
Here {$\bf x(t_i)$} and {$\bf c(t_i)$} are the measured and calculated
position vectors and  $\sigma$ the measurement uncertainties 
as a function of time, $m$ is the number of observed data points
and $n$ the number of free parameters. We have used $n=2$ since
$v_{z}$ and $s_{z}$ are undetermined
 and currently the dominant source of uncertainty (see \ref{sec4.3}).
The pairs $v_{z}$ and $s_{z}$ and $-v_{z}$ and $-s_{z}$ result 
in the same projected orbits and $\chi^2$ values.
Using the orbital data point that is closest to 
our reference position (see Tab.\ref{t02}) we synchronized the 
densely sampled calculated orbit with the measurements.
In Fig.\ref{fig07} we show diagrams for the type for simulations
described above.
For the stars S1, S2, and S8 the resulting $\chi^2$ values are shown in the
$v_{z}$-$s_{z}$-plane in Fig.~\ref{fig08}, ~\ref{fig10}, and ~\ref{fig12}.
In those diagrams we can in general distinguish between three areas 
labeled A, B, and C in Fig.\ref{fig07}: 

\begin{itemize}   
\item[A]
At small separations from Sgr~A$^*$ ($s_z<5~mpc$) the calculated 
orbits have acceleration values which are well above what is measured. 
The corresponding $\chi^2_A$ values are highest. 
These orbits can clearly be excluded.
\item[B]
At large separations from Sgr~A$^*$ ($s_z>10~mpc$) or large line of
sight velocities ($|v_z| > 2000~km/s$) the orbits result in
linear trajectories over the time interval from 1992 to 2000.
The accelerations are too small.
These orbital solutions can be excluded as well. 
Large $\chi^2_B$ values in that region are due to the 
measurement uncertainties as well as a mismatch with respect to a 
straight line. This mismatch is due to the curvature in the 
measured orbital section.
\item[C]
Finally, there is an area in the $v_{z}$-$s_{z}$-plane in which the $\chi^2$
values are lowest and correspond to acceptable orbital solutions
with curvatures similar to what is measured.
These minimum fit errors $\chi^2_C$ are only dominated by the 
scatter in the data.
The difference between $\chi^2_B$ and $\chi^2_C$ is 
a measure of the true $\chi^2$ deviation of the measured 
curved orbital section from a simple linear trajectory - not 
contaminated by the scatter in the data.
\end{itemize}

In Fig.~\ref{fig07} we show the results of 
orbit calculations applied to simulated data for stars similar to S2.
Compared to the available measurements these data have a similar 
sampling but are noise free with respect to the calculated orbits 
from which they have been drawn.
The calculations show that the shape of the 
$\chi^2$ minima depends on the orbital section 
for which measurements are available.
Towards larger velocities and line of sight separations from SgrA*,
i.e. lower orbital curvatures, it becomes increasingly difficult 
to distinguish between possible orbital solutions.
The location of the minimum $\chi^2$ values are smeared out towards 
this region.
For a less curved section the line of sight separation 
can be higher to result in a similarly curved orbit section 
over the same amount of time and hence minimum $\chi^2$ 
values at a higher velocity.
Despite of this effect the simulations also show that 
a common intersection (marked with a filled circle in Fig.~\ref{fig07})
of the regions of minimum $\chi^2$ values
remains at the correct $v_{z}$- and $s_{z}$-values with 
which the stellar orbits are launched at the corresponding epoch
assumed for the simulations - excepting of course the ambiguity in 
the sign of those quantities (see section \ref{sec1.1}). 
How deep and close the absolute minimum of the $\chi^2$ values is
with respect to this location depends on the resolution (sampling in 
the $v_{z}$-$s_{z}$-plane) of the calculation, the signal to noise,
and sampling of the observations.

To get a clear measure of the true $\chi^2_*$ deviation of the measured 
curved orbital section from a simple linear trajectory - not 
contaminated by the scatter in the data -
we corrected for both the SHARP/NTT and the NIRC/Keck data 
the $\chi^2$ values by the corresponding minimum $\chi^2_C$ values
(see before).
\begin{equation}
\label{eq040}
\chi_*^2= \chi^2 - \chi^2_C
\end{equation}

We then combined both data sets in a maximum 
likelihood (ML) analysis via:

\begin{equation}
\label{eq000}
log(ML) =  - \chi_{SHARP/NTT}^2/2 - \chi_{NIRC/Keck}^2/2.
\end{equation}
The results are shown on the right hand site panels of
Fig.~\ref{fig08} ,~\ref{fig10}, and~\ref{fig12} and discussed in the following 
section.

\subsection{
\label{sec4.2}
The central high velocity stars} 
\normalsize 
\noindent

In the previous section \ref{sec4.1} we presented a general discussion
of the procedure we use to match the data with Keplerian orbits.
We now discuss detailed results for the individual stars
obtained from the $\chi^2$ fits in the $v_{z}$-$s_{z}$-planes
and present characteristic orbits.
We show that the high velocity stars S2, and most likely S1 and S8 as well
are on bound, inclined ($60^o<i<80^o$), and eccentric ($0.4\leq$e$\leq1.0$) 
orbits around a central, dark mass. For these 3 stars we list the 
3-dimensional positions and velocities at the time averaged epoch of 
the current SHARP/NTT data set (1996.5) and the orbital elements 
derived from the combined SHARP/NTT and NIRC/Keck 
data sets in Tab.\ref{t06}.
Our analysis also demonstrates that right now it is only
the combination of the two sets that actually allows a first derivation
of orbital elements.   

\subsubsection{
\label{sec4.2.1}
Shapes of the orbits}
\normalsize 

\noindent
{\bf S1 and S2:}
From the $v_{z}$-$s_{z}$-planes shown in Fig.~\ref{fig08} and Fig.~\ref{fig10}
two classes of orbits can clearly be excluded. These are those
with high curvature, corresponding to small distances to the
plane of the sky in which SgrA* is located, as well as those orbits
with small curvatures ($< 1mas/yr^{-2}$) and high eccentricities ($e>1.0$).
The orbital calculations reveal a well defined single $\chi^2$ minimum.
For star S1 about 80\% of the $log(ML)$ values within the
1~$\sigma$ contour in Fig.~\ref{fig08}d correspond to 
eccentricities of $e\le 1.0$ and the 
separation from the SgrA* plane of the sky is about $\sim$7~$mpc$.
For star S2 {\it all} $log(ML)$ values within the 1~$\sigma$ contours in
Fig.~\ref{fig10} b) and d) are consistent with bound orbits, i.e. e$<$1.0.
Here both data sets (SHARP/NTT and NIRC/Keck) indicated
separations from the SgrA* plane of the sky of $\sim$6~$mpc$
and a line of sight velocity in the range of $\pm$500$km/s$.
For both data sets and sources characteristic orbital solutions are shown in 
Fig.~\ref{fig09} and Fig.~\ref{fig11}. 
For both stars the eccentricities are most likely in the range of 
$0.4\le e < 1.0$.
For larger and smaller values of 
$s_z$ the eccentricities and half axes become correspondingly larger and 
smaller (Tab.~\ref{t06}).
For S2 the orbital elements listed in Tab.\ref{t06} are defined best.
For S1 about 20\% of the orbital fits obtained from the possible 
$v_{z}$-$s_{z}$ points (Fig.\ref{fig08}d) result in large semi-major 
axes and high eccentricities.

\noindent
{\bf S8:}
The $v_{z}$-$s_{z}$-planes are shown in Fig.~\ref{fig12} and 
for both data sets we show characteristic Keplerian orbits 
in Fig.~\ref{fig13} (see also Tab.~\ref{t06}).
From Fig.~\ref{fig12}, however, it is evident that there is 
a clear mismatch between the measured curvature values 
and those indicated by the  $\chi^2$  minima.
These minima show that pure Keplerian orbits 
result in a curvature of 0.6 to 0.8 $mas~yr^{-2}$
rather than about 3 $mas~yr^{-2}$ as obtained by the NTT and Keck 
proper motion experiments (see Tab.\ref{t01}).
This mismatch corresponds to a 3-4$\sigma$ deviation 
from the measured value. The correspondence would be better for central masses
above 3$\times$10$^{6}$\solm. However, already the lower enclosed 
mass limit derived from the accelerations of star S8  
represents a 3-4$\sigma$ deviation
from the values obtained via Jeans modeling and other mass estimations based on
proper motions and Doppler velocities (see Tab.\ref{t05} and Fig.\ref{fig06}
in this paper and Tab.5 and Fig.17 by Genzel et al. 2000).
\\
The orbital elements for S8 listed in Tab.\ref{t06} correspond to
the best fits to the data shown in Fig.\ref{fig01} in this paper and
 and Fig.\ref{fig01} in Ghez et al. (2000) and reproduce the observed
time averaged positions and velocities but not the curvatures
(see discussion in section \ref{sec4.2.2}).
Orbits with large curvatures can clearly be excluded.
The most probable eccentricities are just below  $e\sim1.0$.
The proper motion velocity is too large for orbits with apoastron
positions, i.e. regions of higher curvature closer to the present 
location of the star. This is a strong indication for the fact 
that the observed amount of curvature is not solely due to orbital motion.
If this result is confirmed by further measurements 
consequences are that S8 cannot be used to 
pinpoint the location of SgrA* (see section \ref{sec2.2}).

{\bf Influence of the position and amount of the central mass:}
The errors of the orbital elements in Tab.\ref{t06} have been 
derived from the uncertainties of the 3D-positions and velocities
listed below.
 Since the possible range of $v_{z}$ and  $s_{z}$ that results 
 from our fit is large compared to
 the measurement uncertainties of the proper motion velocities
 and positions (see Tab.\ref{t02})
 the resulting uncertainties on the orbital elements are much 
 smaller and well covered by their errors listed in Tab.\ref{t06}.
\\
 The $\pm$30~$mas$  (Menten et al. 1998) uncertainty 
 of the position of SgrA* - which we assume to be
 associated with the central mass - is comparable to the uncertainty 
 of the line of sight separation $s_{z}$.
 It amounts, however, to only less than about 1/8 of the 3-D 
 separation of the stars from SgrA*. 
A simultaneous variation of the amount and position of the central mass 
within the $\pm$30~$mas$ and the (2.6-3.3)$\times$10$^6$\solm~ intervals 
shows that the orbital elements in Tab.\ref{t06} represent a solution 
at the global $\chi^2$ minimum of the orbital fits to the measured data.
 We find that such a variation of the position and amount
 of the central mass causes changes in the eccentricities and the 
 semi-major axes that are well covered by the errors of the orbital 
 elements given in Tab.\ref{t06}.
 Therefore the main result - that the central stars are on 
 fairly inclined and eccentric orbits - is independent of the
 variation of the involved quantities within their errors.

\subsubsection{
\label{sec4.2.2}
What causes the acceleration of S8~?}
\normalsize 
\noindent

In the previous section we have shown that the orbital curvature observed 
by both proper motion experiments is too large for being solely due to
Keplerian motion.
We now discuss a variety of reasons that could explain the
observed acceleration of the star S8.

{\it Stellar scattering:}
The curvature of S8's orbit corresponds to a deviation 
from a straight line by an angle $\psi$. 
If this is caused by a scattering star
of mass $m$, then its distance $r_s$ from S8 is given by

\begin{equation}
\label{eqs1}
r_s \sim 2G (m_{S8}+m)/(v_{\infty}^2 \psi) 
\end{equation}

(Binney \& Tremaine 1994), where
$v_{\infty}$ is the relative velocity at infinity between the two stars.
The probability of such a scattering event occurring during the time 
$\Delta t$ of the monitoring campaign is 

\begin{equation}
\label{eqs2}
P = \pi r_s^2 n v_{\infty} \Delta t 
  \approx 4 \pi G^2 n (m_{S8}+m)^2 \Delta t/(v_{\infty}^3 \psi^2),
\end{equation}

where $n$ is the stellar number density.
Here $v_{\infty}\sim$$\sigma_{central}$$\sim$500 km/s  
corresponds to the velocity 
difference of both stars at large separations.
The mass of S8 is assumed to be 
$m_{S8}\sim15-20$\solm~(Eckart, Genzel, Ott 1999, Genzel et al. 1998).
This implies  $m'$=$(m_{S8}+m)\sim$20\solm~ for $m$$\le$1\solm. 
With the central stellar mass density given by (Genzel et al. 1998, 2000) 
we assume that the stellar number density is of the order of 
$n$$\sim$10$^6$ pc$^{-3}$.
From the acceleration values in Tab.\ref{t01} we derive an observed
scattering angle $\psi$ of the order of 20 degrees for S8.
Equation \ref{eqs2} can then be written as

\begin{equation}
\label{eqs3}
P \approx 5 \times 10^{-8} \times 
     n[10^6 pc^{-3}] (m'[20M_{\odot}])^2 (v_{\infty}[500 km/s])^{-3}
\end{equation}

This shows that even if the stellar number density is higher by a 
few orders of magnitude due to a stellar cusp or if $v_{\infty}$ 
varies by a few 100~km/s the scattering probability is always very low.

{\it Flux density of neighboring stars:}
A K=15-17 background or foreground star close to the current line of 
sight toward S8 could also be responsible for a positional shift
that gives rise to the observed apparent acceleration. 
However, S8 has moved by about 160~$mas$ over the past 8 years.
At a wavelength of 2$\mu$m this corresponds 
to the angular resolving power of the NTT and about 3 times the 
resolving power of the Keck telescope. 
Such a star near S8 has not yet been reported
but - if present and not strongly variable - should be detected soon. 
If the S8 acceleration
is due to such a star the S8 trajectory should straighten again
in the near future.

{\it Alternatives: } 
If other observational biases (e.g. misalignments in position
or position angle) were relevant one would expect even larger 
variations in proper motions at increasing projected separations 
from the center.
These variations are not observed in both independent data sets.
Also a systematical underestimation of the enclosed mass from proper 
motions and radial velocities is not likely. See detailed discussions in
Genzel et al. (2000).
A lensing event can also be excluded as a straightforward explanation 
for the observed acceleration. For stars as bright as S8 such events
are very unlikely and result in a flux density increase over a 
period of approximately 1 year (Alexander \&  Loeb 2001, 
Alexander \& Sternberg 1999).
Within less than about 0.5 magnitudes S8 was constant in 
flux density over the past 8 year.

{\it As a conclusion} the acceleration of S8 that has been detected 
in both the SHARP/NTT and the NIRC/Keck experiment is either due to 
a flux contamination of an unrelated object along the same line of 
sight or due to a rare scattering event in the dense environment 
of the central stellar cluster.

\subsubsection{
\label{sec4.2.3}
Other central early type stars}
\normalsize 
\noindent

For the  remaining  early type stars of the central Sgr~A* cluster
positions and proper motions are known 
(Genzel et al. 1997, 
Ghez et al. 1998,
Ghez et al. 2000, 
Genzel et al. 2000). 
Orbit calculations show that stars in the 
Sgr~A* cluster with line of sight separations from the 
center of $s_z$$<$30~$mpc$ and line of sight velocities 
$v_z$ smaller than 2 to 3 times the velocity dispersion of the 
central arcsecond will be on bound orbits around the black hole.
A more detailed analysis, however, still awaits a detection of their
orbital curvature and/or their radial velocity.

\subsection{
\label{sec4.3}
Late type stars at small projected separations}
\normalsize 
There are two stars with prominent CO band head absorption that are located at
small projected separations from SgrA* and for which the full 3-dimensional
velocity information is available. 
The corresponding $v_{z}$-$s_{z}$-planes and characteristic orbital 
solutions are shown in Fig.~\ref{fig14}.
It cannot be fully excluded that these stars are at small 
physical distances to the center.  Our orbital analysis, however, shows that 
the current data suggest a likely location outside 
the central 0.3~pc diameter section of the Galactic Center stellar cluster 
which is dominated by the early type He-stars. In the following we discuss the 
results of our orbital analysis for both stars.
 
{\it No.25 - 0.43''E; 0.96''S of SgrA*:} Based on R$\sim$5000
VLT  ISAAC observations Eckart, Ott,\& Genzel (1999) report 
the presence of a late type star with strong 2.3$\mu$m CO band head absorption 
about 1'' south of the center. 
We identify this object with the K=12.4 proper motion star No.25 in Tab.1 
of Genzel et al. (2000) and star S1-5 in Tab.1 by Ghez et al. (1998).
This star is located at a projected separation of
1.05'' about 0.43''E and 0.96''S of SgrA*. This star is approximately 0.3
magnitudes brighter than the overall southern part of the SgrA* cluster 
(containing S9, S10, S11, and a few K$\ge$16 stars just E of S10 and S11).
In a 0.3''-0.5'' seeing under which the VLT data (Eckart, Ott, Genzel 1999) 
were taken the flux density contribution of this star in a 0.6'' NS slit 
is comparable to that of the southern part of the SgrA* cluster. 
A fainter star almost exactly 1.1''S of SgrA*
can be excluded as a possible identification of the late type star, since its
brightness is about 0.3 magnitudes fainter than the individual stars S11 or 
S10 and hence almost a full magnitude fainter than the total of the 
southern part of the SgrA* cluster.
The wavelength calibration of the R$\sim$5000 VLT ISAAC data - as well as 
a comparison to the spectrum of the late type star IRS14SW (see Tab.1 by
Genzel et al. 2000) that fell into the NS oriented slit and was acquired
simultaneously - indicate a line of sight velocity of the star 
0.43''E and 0.96''S of SgrA* of -80$\pm$40 km/s.
The SHARP/NTT proper motion data of this object including the results 
of the 1999 and 2000 observing run are shown in Fig.\ref{fig02}
 and listed in Tab.\ref{t03}.
A comparison of the radial and proper motion velocities indicates that 
this star is on a predominantly tangential orbit in the plane of the sky.

In Fig.\ref{fig14} we investigate the possible orbital solutions that 
lead to bound orbits. The thick (red) continuous and 
dashed lines mark line of 
sight separations from SgrA* for which the eccentricities $e<1$.
About 60\% of the possible current line of sight separations are located
beyond the radius within which the He-stars dominate the emission. 
About 30\% even lie 
beyond the core radius of the central stellar cluster of $\sim0.3~pc$.
If the line of sight separations $s_z$ are of the order of 150 to 200~mpc 
the eccentricities are smaller than unity and the 
semi-major axes of the orbits will be of the same order as $s_z$.
These numbers are lower limits only, since they are derived for simple 
Keplerian orbits under the assumption of a dominant central mass of 
$3\times10^6$\solm. For orbits with eccentricities (as calculated 
for the simple Keplerian case in Fig.\ref{fig14}) closer to $e=1.0$ and 
values for $s_z$$>$$200~mpc$ the orbits will have large (several degrees) 
Newtonian periastron shifts.
The stars reach true physical separations from the center 
of well beyond 1.0~pc for which the mass of the 
stellar cluster starts to dominate. Under 
these conditions bound stellar orbits with line of sight separations 
larger than what is indicated by the thick (red) lines are possible.
\\
\\
{\it S18 - 0.04''W; 0.45''S of SgrA*:}
This star is listed as S18 in Tab.1 of Genzel et al. (2000) but not 
contained in the corresponding list of Ghez et al. (1998). 
Its most recent SHARP/NTT proper motion data are listed in Tab.~\ref{t03}.
Based on deep CO(2-0) 
line absorption Gezari et al. (2000) identify this object as an early 
K-giant which is blue shifted with respect to the Galactic Center 
stellar cluster at about -300 km/s. Of all late type stars in the central 
stellar cluster S18 has the smallest angular separation ($<$0.5'') 
from SgrA* reported to date.
A comparison of the radial and proper motion velocities indicates that 
this star could be on a predominantly radial orbit.
In Fig.\ref{fig14} we investigate the possible 
orbital solutions that lead to bound 
orbits. The thick (red) continuous and 
dashed lines mark line of sight separations from 
SgrA* for which the eccentricities $e<1$.
About 50\% of the possible current line of sight locations are located beyond
the radius of the He-stars cluster and reach out to the core radius of 
the central stellar cluster. For the reasons mentioned above 
bound stellar orbits with line of sight separations 
larger than what is indicated by the red lines are possible.
\\
\\
Highly eccentric orbits like those labeled with '$I$' in Fig.\ref{fig14} 
 bring both late type stars physically too close to the position of SgrA*.
These orbits can be excluded because 
for a black hole mass of $3\times$10$^6$\solm 
~the tidal disruption radius for a giant is 

\begin{equation}
\label{eq7}
R_t \sim 1.2mas \times (R_{*}/10^{12}cm) \times (M_{*}/ M_{\odot})^{1/3}
\end{equation}

(e.g. Frank \& Rees  1976, Binney \& Tremaine 1994), where 
$R_{*}$ and $M_{*}$ are the giant's radius and mass. For orbits with
semi-major axes of $a \sim 20mpc  \sim 0.5''$ and eccentricities of $e >$0.94
every giant will be destroyed on its periastron passage. The coupling between
the orbit and the tides raised on the star will cause deviations from 
a point mass behavior even at separations larger than $R_t$.
Correspondingly this would allow only wider orbits for giants.
Along the same line of arguments equation \ref{eq7} also provides additional 
evidence that the central high velocity stars (the S-stars) are O-stars rather 
than late type giants. 
This identification, however, still awaits spectroscopic confirmation.

\section{
\label{sec5}
SUMMARY AND CONCLUSIONS}
\normalsize 
The combination of the high precision but shorter time scale NIRC/Keck 
data with the lower precision but longer time scale SHARP/NTT data set
allows us to have a first insight into the nature of 
{\it individual} stellar orbits 
as close to the massive black hole at the center of the Milky Way
as currently possible.

We have shown that a statistical correction for geometrical projection
effects allows us to derive an enclosed mass estimate from the
observed accelerations of stars S1 and S2 of 
$M_{acc}=(5 \pm 3) \times 10^6$\solm.
This value is fully consistent with an enclosed mass that 
is flat down to radii of about 8~mpc with a value of about 
$3 \times 10^6$\solm~and mass density of $3.7 \times 10^{12}$\solm$~pc^{-3}$ 
for a core radius of $r_c=5.8~mpc$ as derived from the proper 
motion data (Genzel et al. 2000).
Our most recent data - compared to and combined with 
published data on proper motions and accelerations 
(Ghez et al. 2000,
Genzel et al. 2000,
Eckart et al. 2000,
Ghez et al. 1999,
Ghez et al. 1998) -
show that S2 - and most likely S1 and S8 as well - are on 
orbits around a central, dark, and massive object coincident with 
the position of the radio source SgrA*.
The stars are on bound fairly inclined ($60^o<i<80^o$) 
and eccentric ($0.4<e<0.95$) orbits. 
In Fig.\ref{fig15} we show for star S2 right ascension and declination 
as a function of time as predicted from our analysis of the combined data set.
The analysis is in agreement with the result by
Genzel et al. (2000) that the central stars are preferentially on 
 eccentric orbits.
This statement also holds for S8 - just 
taking into account its velocity at the current projected position - although
it is likely that not the entire amount of its measured acceleration 
is due to orbital motion.
\\
\\
{\it Properties of a possible cusp:}
The current kinematical data on individual central stars make it difficult
to accommodate a high density cusp with a stellar density law as steep as
$n(r) \sim r^{-7/4}$. In section \ref{sec3.2.2} we have shown that for 
steep cusps the majority of the projected mass estimates is much closer
to the true value than in the case of for instance a constant density 
distribution. For three stars randomly picked out of a cluster following 
a $r^{-7/4}$ density law we would at least for two stars expect that the 
mass estimate derived from their accelerations accounts for $\sim$70\% 
of the enclosed mass. With S8 and possibly S2 this is just barely the case
for the Galactic Center. 
The presence of a pronounced stellar cusp within the central few 100 mpc
would make that area collisionly dominated and most of the stars would 
be expected on parabolic orbits (Alexander 1999).
Our orbit calculations show that the current data of both the SHARP/NTT and 
NIRC/Keck experiment clearly indicate  
that the central stars are on bound and not on parabolic orbits. 
This weakens the case for a {\it strong} cusp and 
suggests that the stellar 
density is not high enough to provide a collisionally dominated environment.
\\
\\
Due to the limited number of detected stars a 
minimum radius of 10~mpc (0.25'') is currently used for the 
determination of the mass and mass density from proper motion measurements.
The $\alpha$=5 Plummer like  model of a dark cluster 
results in a core radius of such a hypothetical cluster of
r$_{core}$=5.8~mpc (0.15'') and corresponding central density
of 3.7$\times$10$^{12}$\solm~pc$^{-3}$.
The statistical correction for geometrical projection effects 
presented in section ~\ref{sec3.2} shows that the acceleration 
data of the two closest stars S1 and S2 are in agreement with 
such a compact mass.
Accepting the correction for the
most likely volume that contains S1 and S2 - the data would even be fully
consistent with a core radius of r$_{core}$=4.2~mpc (0.17'') and 
corresponding central density of 10$^{13}$\solm~pc$^{-3}$.
If the trajectory of S2 remains consistent with an orbit around a
compact mass of 3.0$\times$10$^6$\solm ~~the lower limit for the
mass density that can be derived from this stellar orbit
may be as large as 2.4$\times$10$^{14}$\solm~pc$^{-3}$.
In this case the collapse life time of a hypothetical cluster of dark
mass would shrink to only a few 10$^6$ years (Maoz 1998).
\\
\\
Two late type stars for which proper motion and radial velocities are known, 
and that are at projected separations of about 0.5'' and 1.0'' from SgrA* 
are very likely at larger physical distances from the center and part 
of the larger scale central stellar cluster with a core radius of 
approximately $0.3~pc$.
Our analysis indicates that the strong curvature of the 
available orbital section of star S8 is possibly due to 
either a flux contamination by an unrelated object along the same line of 
sight or a result of a rare scattering event in the dense environment 
of the central stellar cluster.
\\
\\
 The analysis of stellar orbits 
(Ghez et al. 2000, Genzel et al. 2000, Eckart et al. 2000, and this paper)
clearly supports the presence of a 
compact dark mass at the position of SgrA*.
The VLBI maser nucleus of NGC~4258 (Greenhill et al. 1995,
Myoshi et al. 1995) and the dark mass at the center of the 
Milky Way are currently 
the best and most compelling cases for the existence of
super-massive nuclear black holes (Maoz 1998).
\\
\\
{\it Future developments:}
Future observations at higher sensitivity and angular resolution
will allow us to find and track the motion of stars that are even 
closer, i.e. $<4~mpc$, to the center than those currently accessible. 
Orbital time scales at the resolution limit of the VLT, LBT, or Keck 
interferometer could be in the range of a few months. 
A detection of a relativistic or 
Newtonian periastron shift would ultimately result in a direct determination 
of the compactness of the enclosed central mass 
(Rubilar \& Eckart 2000; see also Fragile \& Mathews 2000).
Measurements of the prograde relativistic or the retrograde 
Newtonian orbital periastron shifts are within reach of current and 
upcoming instrumentation.
 
\acknowledgements
We are thankful to the ESO Director General and his staff to let us bring
the SHARP camera to the NTT in 1999 and 2000.
We thank Reiner Hofmann and Klaus Bickert for their help with SHARP at the NTT.
We are also grateful to the NTT  and La Silla team for their interest 
and technical support of the SHARP camera.

\clearpage
\newpage
\large
\begin{center}
{\bf TABLES}
\end{center}
\normalsize

\begin{table}[!htb]
\caption{
\label{t01}
Accelerations}
\begin{center}
\begin{tabular}{rrrrrrrrr}\hline \hline
& & & SHARP & & Keck & \\ 
& d$_{Sgr~A*}$ & $a_{\alpha}$ & $a_{\delta}$  & a$_{1996.5}$ 
& d$_{Sgr~A*}$ & a$_{1995.53}$ \\
\\ \hline 
source: 
S1 & 3.10$\pm$0.12 &+1.5 $\pm$ 1.7 & -3.5 $\pm$ 1.8 & 3.8$\pm$2.4 & 3.44$\pm$0.03 & 2.4$\pm$0.7 \\
S2 & 5.37$\pm$0.12 & +1.0 $\pm$ 0.7 & -2.1 $\pm$ 0.6 & 2.3$\pm$0.9 & 4.61$\pm$0.04 & 5.4$\pm$0.3 \\
S8 & 13.7$\pm$0.12  &-3.3 $\pm$ 0.7 & -0.3 $\pm$ 0.9 & 3.3$\pm$1.1 &14.64$\pm$0.04 & 3.2$\pm$0.5 \\
S7 & 24.5$\pm$0.12   & & & $<$3.0  &               &              \\
S10& 17.8$\pm$0.12   & & & $<$4.0  &               &              \\
S11& 21.9$\pm$0.12   & & & $<$3.0  &               &              \\
S18& 17.5$\pm$0.12   & & & $<$3.0  &               &              \\
No.25& 40.7$\pm$0.15 & & & $<$3.0  &               &              \\
\hline \hline
\end{tabular}
\end{center}
Projected separations $d$ from SgrA* in $mpc$ and stellar accelerations 
$a$ in units of $mas~yr^{-2}$ $(1.203\times 10^{-3}m~s^{-2}$)
for the SHARP/NTT (1996.50) and NIRC/Keck (1995.53) data.
For the SHARP/NTT data we also give R.A. and Dec. acceleration 
components $a_{\alpha}$ and $a_{\delta}$ from which the
total orbital acceleration
$a_{total}$=($a_{\alpha}^2$+$a_{\delta}^2$)$^{1/2}$
has been derived.
\end{table}

\begin{table}[!htb]
\caption{
\label{t02}
Positions}
\begin{center}
\begin{tabular}{rrrrrrrrr}\hline \hline
& SHARP & & Keck & \\
epoch & 1996.5 & & 1997.6 & \\
& $\Delta$$\alpha$ & $\Delta$$\delta$ 
& $\Delta$$\alpha$ & $\Delta$$\delta$ \\
\\ \hline 
source: 
S1    & -0.080 & -0.020 & -0.083 & -0.037 \\
S2    & -0.010 & +0.142 & -0.012 & +0.121 \\
S8    & +0.320 & -0.170 & -0.338 & -0.189 \\
S18   & -0.040 & -0.450 & -0.037 & -0.443 \\
No.25 & +0.430 & -0.960 &    -   &    -   \\
\hline \hline
\end{tabular}
\end{center}
Positions relative to SgrA* in $arcsec$ for the
mean center epochs of 1996.5 (SHARP) and 1997.6 (Keck) as derived from 
our own data and the data presented by Ghez et al. (2000).
The assumed relative 1$\sigma$ errors are 10~mas (SHARP) and 5~mas (Keck,
Ghez et al. 2000), respectively. 
For star No.25 and S18 data are taken from Genzel et al. (2000).
Keck data for S18 are taken from Gezari et al. (2000). 
\end{table}
 
\begin{table}[!htb]
\caption{
\label{t03}
Proper Motion Velocities}
\begin{center}
\begin{tabular}{rrrrrrrrr}\hline \hline
& SHARP & & Keck \\ 
& v$_{\alpha}$ & v$_{\delta}$ & v$_{\alpha}$ & v$_{\delta}$ \\ 
\\ \hline 
S1 & +541$\pm$60 &-1460$\pm$60 & +568$\pm$60 &-1528$\pm$60 \\
S2 & -290$\pm$60 & -694$\pm$60 & -271$\pm$60 & -733$\pm$60\\
S8 & +535$\pm$60 & -552$\pm$60 & +472$\pm$60 & -558$\pm$60 \\
S18& +3  $\pm$60 &  146$\pm$60 & -  & - \\  
No.25 & -227$\pm$60 &  14$\pm$60 & -  & - \\  
\hline \hline
\end{tabular}
\end{center}
The velocities  in $km~s^{-1}$ were obtained for 
epochs 1996.5 (SHARP) and 1997.6 (Keck) 
via interpolation of the measured velocities for epochs 1996.5 (SHARP)
and 1995.4 (Keck) with the measured projected accelerations in R.A. and Dec.
(see Tab~\ref{t01}).
For S18 a radial velocity of $v_z=$-300~km/s is given by Gezari et al. (2000), 
and for No.25 a radial velocity of -80~km/s is derived from ISAAC VLT data
(this paper and Eckart, Ott, Genzel, 1999).
For errors see caption to Tab.~\ref{tapp01}.
\end{table}

\begin{table}[!htb]
\caption{
\label{t04}
Correction Factors}
\begin{center}
\begin{tabular}{rrrrrrrrrrr}\hline \hline
&$r_V$  & S1 & & S2& & S8& \\
&[arcsec]&cos($\theta$)& cos($\theta$)$^{-3}$& 
cos($\theta$)& cos($\theta$)$^{-3}$& 
cos($\theta$)& cos($\theta$)$^{-3}$ \\
\\ \hline 
Keck & & & & & & & \\
&0.3 &0.54 $\pm$0.17 &6.44$\pm$5.1  &0.67 $\pm$0.17 &3.4 $\pm$2.2 &1.0 &1.0 \\
&0.4 &0.42 $\pm$0.15 &13  $\pm$12   &0.54 $\pm$0.17 &6.4 $\pm$5.0 &0.99 $\pm$0.07 
&1.02 $\pm$0.02\\
&0.5 &0.35 $\pm$0.13 &24$\pm$22     &0.45 $\pm$0.16 &11  $\pm$9 &0.93 $\pm$0.07 
&1.26 $\pm$0.23\\
\hline
& & & & & & & \\
SHARP & & & & & & & \\
&0.3 &0.50 $\pm$0.17 &8.3 $\pm$6.9 &0.73 $\pm$0.16 &2.5 $\pm$1.4 &1.0 &1.0 \\
&0.4 &0.39 $\pm$0.14 &17   $\pm$15 &0.61 $\pm$0.18 &4.5 $\pm$3.3 &0.97 $\pm$0.02 
&1.08 $\pm$0.07\\
&0.5 &0.32 $\pm$0.12 &32   $\pm$29 &0.51 $\pm$0.17 &7.5 $\pm$6.1 &0.90 $\pm$0.08 
&1.36 $\pm$0.32\\
\hline \hline
\end{tabular}
\end{center}
The geometrical correction factors have been calculated for a constant
number distribution $n(r)$ of stars within the central volume 
in the potential of a central compact mass as indicated by current 
observations (see equation (33) and Tab.5 by Genzel et al. 2000).
\end{table}

\begin{table}[!htb]
\caption{
\label{t05}
Corrected Mass Estimates}
\begin{center}
\begin{tabular}{rrrrrrrrrrr}\hline \hline
& S1 & & S2& & S8& \\
&radius & mass & radius & mass & radius & mass \\
&[mpc] & [10$^6$\solm] & [mpc] & [10$^6$\solm] & [mpc] & [10$^6$\solm] \\
\\ \hline 
SHARP 
&8.0 $\pm$2 &6.4$\pm$3.8 & 9.0$\pm$ 1.6 &2.9 $\pm$1.5 & 14.4$\pm$0.8 & 6.2 $\pm$ 1.0 \\
& & & & & & & \\
Keck
&8.1 $\pm$2 &3.7$\pm$2.1 &8.6 $\pm$1.7 & 7.2$\pm$3.8 & 15.2$\pm$0.5 &6.7 $\pm$ 0.8 \\
\hline \hline
\end{tabular}
\end{center}
Here we used the correction factors for a flat stellar distribution
listed in Tab.\ref{t04}. 
\end{table}

\begin{table}[!htb]
\caption{
\label{t06}
Orbital parameters}
\begin{center}
\begin{tabular}{lllllllll}\hline \hline
\\
source               & S1                          &  S2                    &  S8      
\\
\\ \hline
\\
$i$~~[$^{\circ}$]    & 60$^{+5}_{-5}$              & 70$^{+6}_{-17}$        & 0$^{+70}_{-70}$  \\
$\omega$~[$^{\circ}$]& 190$^{+120}_{-10}$          & 190$^{+10}_{-10}$      & -55$^{+32}_{-0}$ \\
$\Omega$~[$^{\circ}$]&-14$^{+7}_{-12}$             & 23$^{+27}_{-11}$       & $\sim$-65        \\
$e$                  & 0.6$^{+0.30(*)}_{-0.30}$& 0.8$^{+0.15}_{-0.40}$  & 0.95$^{+0}_{-0.12}$\\
$a$~[$mpc$]          & 18$^{+18 (*)}_{-8}$      & 5.6$^{+0.7}_{-1.1}$    & 10$^{+48}_{-0}$   \\
$T$~[$yrs$]          & 2097$^{+281(*)}_{-96}$ & 2002.6$^{+6.8}_{-2.2}$ & 2043$^{+1060}_{-0}$\\ 
$P$~[$yrs$]          & 100$^{+184(*)}_{-54}$  & 19.4$^{+7.4}_{-3.0}$   & 57$^{+1060}_{-0}$ 
\\
\\ 
\hline
\\  
 $s_{\alpha}$~~[$mpc$] & -3.02$\pm$0.19& -0.38$\pm$0.19 & +12.10$\pm$0.19 \\
 $s_{\delta}$~~[$mpc$] & -0.76$\pm$0.19& +5.38$\pm$0.19 & -6.43$\pm$0.19 \\
 $s_{z}$~~[$mpc$]      & +7$\pm$1      & +6$\pm$1       & $<$4 \\ 
 $v_{\alpha}$ [$km/s$] & +541$\pm$60   &  -290$\pm$60   & +535$\pm$60  \\
 $v_{\delta}$ [$km/s$] & -1460$\pm$60  &   -694$\pm$60  &  -552$\pm$60  \\ 
 $v_{z}$ [$km/s$]      & -400$^{+800}_{-300}$ & 0$\pm$500 &  0$\pm$1000 \\
 \\  
\hline \hline
\end{tabular}
\end{center}
{\bf Top:}
Orbital elements for the stars S1, S2, S8 as derived from the combined 
SHARP/NTT and NIRC/Keck data sets (see Tab.\ref{t06}). 
We listed the orbital elements for the orbits labeled 'I' in 
Figs.\ref{fig08}-\ref{fig13}. The uncertainties correspond to the variation
of orbital elements within the $1 \sigma$ contour
of Figs.\ref{fig09}, \ref{fig11}, and \ref{fig13}.
The orbital elements for S2 are defined best.
 For S1 we have listed the orbital parameters for the $\sim$80\% of the 
fit areas within the  $1 \sigma$ contours
with eccentricities below $e=0.9$. 
In the remaining $\sim$20\% of the cases the quantities marked with an
asterisk may become significantly larger. These orbits for S1
may even be not bound.
\\
We listed the inclination $i$, the sky position angle $\omega$ 
of the periastron position, the sky position angle $\Omega$ of the 
northern most knot (transition between
 dashed and solid sections of the orbits in Figs.\ref{fig08}, \ref{fig10}, and
 \ref{fig12}), as well as the eccentricity $e$ and semi-major axes $a$ of the
 deprojected orbits, and finally the year $T$ of the periastron transition
and the orbital period $P$. 
Future spectroscopic measurements have to determine the sign of the 
inclination $i$ as well as  which of the knots is the ascending one.
The errors are derived from the uncertainties of the 3D-positions and velocities
listed below.
\\
{\bf Bottom:}
3D-positions and velocities at the
mean center epochs of 1996.5 as derived for the SHARP/NTT data. 
The range of line of sight separations
 $s_{z}$  and velocities $v_{z}$ result from the combined 
SHARP/NTT and NIRC/Keck data sets.
The  $s_{z}$ and $v_{z}$ pairs are uncertain by a common 
sign factor of $\pm1$.
\end{table}

\clearpage
\newpage
\large
\begin{center}
{\bf FIGURE CAPTIONS}
\end{center}
\normalsize
\figcaption[]{
\label{fig01} 
Positions of the stars S1, S2, S8, S7, S10, and S11 
relative to the position of SgrA*
as a function of the SHARP/NTT observing epoch from 1992 till 2000. 
For the stars  S1, S2, S8 we show parabolic variations of the positions
that correspond to the data presented in the text and in Tab.\ref{t01}.
For the other stars for which only upper limits 
of the curvature were derived we show linear fits.
The declination velocity plot has been shifted by +300~mas.
}
\figcaption[]{
\label{fig02} 
Positions of the two late type stars No.25 (0.43''E; 0.96''S)
in Tab.1 by Genzel et al. (2000) 
and S18 (0.04''W; 0.45''S) relative to the position of SgrA*
as a function of the SHARP/NTT observing epoch from 1992 till 2000. 
The data shows no indication for significant orbital curvature
above a value of about $3~mas~yr^{-2}$.
The declination velocity plot has been shifted by +300~mas.
}

\figcaption[]{
\label{fig03} 
The position of the central dark mass derived from the SHARP/NTT and 
NIRC/Keck acceleration data of S1 and S2.
For the stars S1 and S2 the measured acceleration data corresponds well to
curvature values expected from orbital motion around a central dark mass.
\\
The grey shading is proportional to the probability of the location 
of the central dark mass, i.e. the maximum likelihood score given in the text. 
The smaller probability plot for the combined SHARP/NTT-NIRC/Keck 
acceleration data has been inserted into the probability plot derived 
from the SHARP/NTT data alone.
The central filled dashed circle marks the radio position 
of SgrA* and corresponding uncertainties of $\pm$30~mas.
The thin contour lines indicate limit at which 
the $log(ML)$-score drops by a factor of 0.5 below its maximum.
The $\alpha$ and $\delta$ coordinates of the central mass
can be read off from the projections of the thin contour lines at the
68.4\% and from the thick contour line at the 90\% level.
The time averaged positions of the SHARP/NTT and NIRC/Keck data 
as well as the error cones of the acceleration data for 
S1 and S2 are indicated by different symbols explained in the figure legend.
For the NIRC/Keck we have taken the error cone widths
used in Fig.2 of Ghez et al. (2000) rather than those 
implied by the values in their Tab.1. 
The positional uncertainties for the SHARP and NIRC positions have been
included in the grey scale plot and the width of the cones.
}

\figcaption[]{
\label{fig04} 
Relevant quantities to correct the measured
accelerations of stars near the Galactic center for projection effects.
}
\figcaption[]{
\label{fig05} 
Enclosed masses as a function of projected and true 
separation from SgrA$^*$ derived from observed stellar acceleration data
for different distributions $\rho(r)$ of the central compact dark mass and the
surrounding stars $n(r)$ within the central volume of 0.4''
(15~mpc) radius.
The top three images (a-c) show the lower limits of the enclosed mass from 
observed accelerations.
The bottom three images (d-f) show the enclosed mass from 
accelerations statistically corrected for projection effects.
The left two graphs assume a 3$\times$10$^6$\solm~point mass and 
a constant number density of stars. The two graphs in the middle assume 
a 3$\times$10$^6$\solm~point mass and a $r^{-7/4}$ number 
density of stars.
The two graphs to the right assume an extended central mass distribution
$\rho(r)$ and a constant number density of stars.
Contour lines are at 5, 10, 20 and 50\% of the peak density of
data points.
\\
In all panels we also show the mass estimates obtained via Jeans modeling 
(indicated by the open box) and other mass estimations (grey shaded box)
based on proper motions and Doppler velocities 
(see Fig.17 by Genzel et al. 2000).
}
\figcaption[]{
\label{fig06} 
Enclosed masses calculated from observed orbital accelerations.
\\
a) Lower limits for the enclosed mass estimates 
from the projected accelerations.
The data points are labeled with source names and origin of the measurements.
The sizes of the error bars are dominated by the uncertainties in the 
determination of the accelerations.
We plotted the SHARP/NTT data as derived for the mean observing epoch.
In the case of the NIRC/Keck data we used the quantities given 
in Table 1 of Ghez et al. (2000).
\\
b) Mass estimates with the statistical correction for geometrical projection
effects described in the text. 
The data points represent the estimates for a volume radius (see text section
\ref{sec3.2}) of 0.4'' (15~mpc). 
Data points and labels for the individual sources 
have been enclosed in a dotted line. 
The area shaded in light grey includes the enclosed mass estimates corrected 
for central volume radii ranging between 0.3'' (11~mpc) and 0.5'' (19~mpc).
The corrected mass estimates have been derived for a constant
number distribution $n(r)$ of stars within the central volume 
in the potential of a central compact mass consistent with
current observations (see equation (33) and Tab.5 by Genzel et al. 2000).
See text for further explanation.
\\
In both panels we also show the mass estimates obtained via Jeans modeling 
(indicated by the open box) and other mass estimations (grey shaded box)
based on proper motions and Doppler velocities 
(see Fig.17 by Genzel et al. 2000).
}
\figcaption[]{
\label{fig07}
Orbital fit calculations applied to simulated data 
on a star similar to S2. 
The calculations were carried out for two data sets of the same orbit.
The results of the calculations are shown on the left, the corresponding 
orbits on the right.
The simulations show that the shape of the distribution of minimum
$\chi^2$ values (inverted grey scale shading and corresponding contour
 lines) is a 
function of the orbit section to which the calculations 
are applied (see text for explanations).
For comparison we have plotted the contour line representation of the results
for the two orbital sections in both $v_{z}$-$s_{z}$-planes on the left.
The filled circle marks the common intersection of the regions of 
minimum $\chi^2$ values at the position of the 'true' current line of sight 
separation and radial velocity of the star - excepting of course the ambiguity 
in the sign of those quantities. 
The asterisk denotes the position of the central mass.
}
\figcaption[]{
\label{fig08} 
Orbital fits for the star S1. 
\\
{\bf a)} The reduced  $\chi^2$ values calculated from the 
1992 to 2000 SHARP/NTT 
measurements and orbits computed from the data in Tab.~\ref{t02},
Tab.~\ref{t03} and the corresponding line of sight 
velocities and separations from SgrA*, as given by the labels of the axes.
The $\chi^2$ values are represented in an inverted grey scale and by the 
labeled thick contour lines (see equation ~\ref{eq04}). 
Thin contour lines represent the eccentricities $e$ of the 
corresponding orbits. All points within the $e=1.0$ contour lines represent
bound orbits. The $e=0.5$ contour is shown as a dashed line.
\\
{\bf b)} The grey scale plot shows the likelihood score 
(see text) derived via the $\chi^2$ values from the
SHARP/NTT and the NIRC/Keck data 
shown in panel a) and c).
A continuous black contour line indicates the 1~$\sigma$ limit
at which the $log(ML)$ score drops by a factor of 0.5 below its
maximum.
Roman numbers label the positions of representative best fit orbits 
within this contour. The projections of the corresponding orbits 
onto the plane of the sky is shown in Fig.~\ref{fig09}. 
The calculated accelerations $a$ are represented by thick contour lines 
with labels in units of $mas~yr^{-2}$.
Thin contour lines represent the orbital eccentricities $e$.
The $a=0.5~mas~yr^{-2}$ and $e=0.5$ contours are shown as dashed lines.
\\
\\
In panels {\bf c)} and {\bf d)} 
we show the corresponding information 
for the NIRC/Keck data (Ghez et al. 2000). 
While the grey scale likelihood plot is the same in panels {\bf b)}
and  {\bf d)} the labeled contour lines represent the eccentricities 
and accelerations obtained from the orbit calculations only using the 
SHARP/NTT and the NIRC/Keck data, respectively.
}
\figcaption[]{
\label{fig09} 
Selected Keplerian orbits with velocities and 
separations from SgrA$^*$ for star S1 as given by the data in 
Tab.~\ref{t02} and Tab.~\ref{t03}.
The orbit labels correspond to those in 
Fig.~\ref{fig08}b and Fig.~\ref{fig08}d.
The orbits represent the range of best fits (within
the 1 $\sigma$ contour line in Fig.~\ref{fig08}; see text)
to the 1992 to 2000 SHARP/NTT data (crosses) and the 1996 to 1999 
NIRC/Keck data (crosses also marked by thin lines on either site of the
orbital tracks; Ghez et al. 2000).
The dashed lines indicate those sections of the orbits that are 
- with respect to the current position of the stars -
on the opposite site of the plane of the sky SgrA* is located in
(see also caption of Tab.\ref{t06}).
The projected position of SgrA* is indicated by an asterisk.
}
\figcaption[]{
\label{fig10} 
Orbital fits for the star S2. See caption of Fig.~\ref{fig08} 
and description in text.
}
\figcaption[]{
\label{fig11} 
Selected Keplerian orbits for S2. See caption of Fig.~\ref{fig09}.
}
\figcaption[]{
\label{fig12} 
Orbital fits for the star S8. See caption of Fig.~\ref{fig08}
and description in text.
The orbital fits reveal a mismatch between the measured 
curvature values of about $3~mas~yr^{-2}$ and those 
resulting from best fit pure Keplerian orbits 
of 0.6 to $0.8~mas~yr^{-2}$. 
}
\figcaption[]{
\label{fig13} 
Selected Keplerian orbits for S8. See caption of Fig.~\ref{fig09}.
}
\figcaption[]{
\label{fig14} 
Possible orbital solutions 
and selected projected orbits for the two late type stars
No.25 (0.43''E; 0.96''S) and S18 (0.04''W; 0.45''S).
For panels {\bf a)}, {\bf c)} and {\bf b)}, {\bf d)} see captions 
of Fig.~\ref{fig08} and Fig.~\ref{fig09}, respectively.
The thick (red/grey) lines indicate the range in line of sight velocity and 
separation from SgrA* for which bound orbits are possible (see comment on 
lower limits in text). For the thick dashed line the velocities and 
separations have to be multiplied by -1. The orbits IV are not shown.
For 0.43''E; 0.96''S orbit IV is close to an EW flip of orbit III 
and for S18 orbit IV is close to an NS flip of orbit III.
}
 
\figcaption[]{
\label{fig15} 
Prediction of position time plots for different possible orbits of the star S2.
For a line of sight separation from the plane of the sky containing 
Sgr~A* of $s_z=$6~mpc the following line of sight velocities have
been used (see section \ref{sec4} and Fig.\ref{fig11}): 
I: $s_v=$0 km/s; II: $s_v=$-500 km/s; III: $s_v=$+500 km/s.
The longest orbital time scales suggested by the current data 
correspond to orbits with 
$s_z=$7~mpc and $a$: $s_v=$-500 km/s or $b$: $s_v=$500 km/s.
The SHARP/NTT data is shown as well.
The declination velocity plot has been shifted by +300~mas.
}
 
\clearpage
\newpage
\begin{figure}[!htp] 
\begin{center}
\begin{tabular}{c}
\psfig{figure=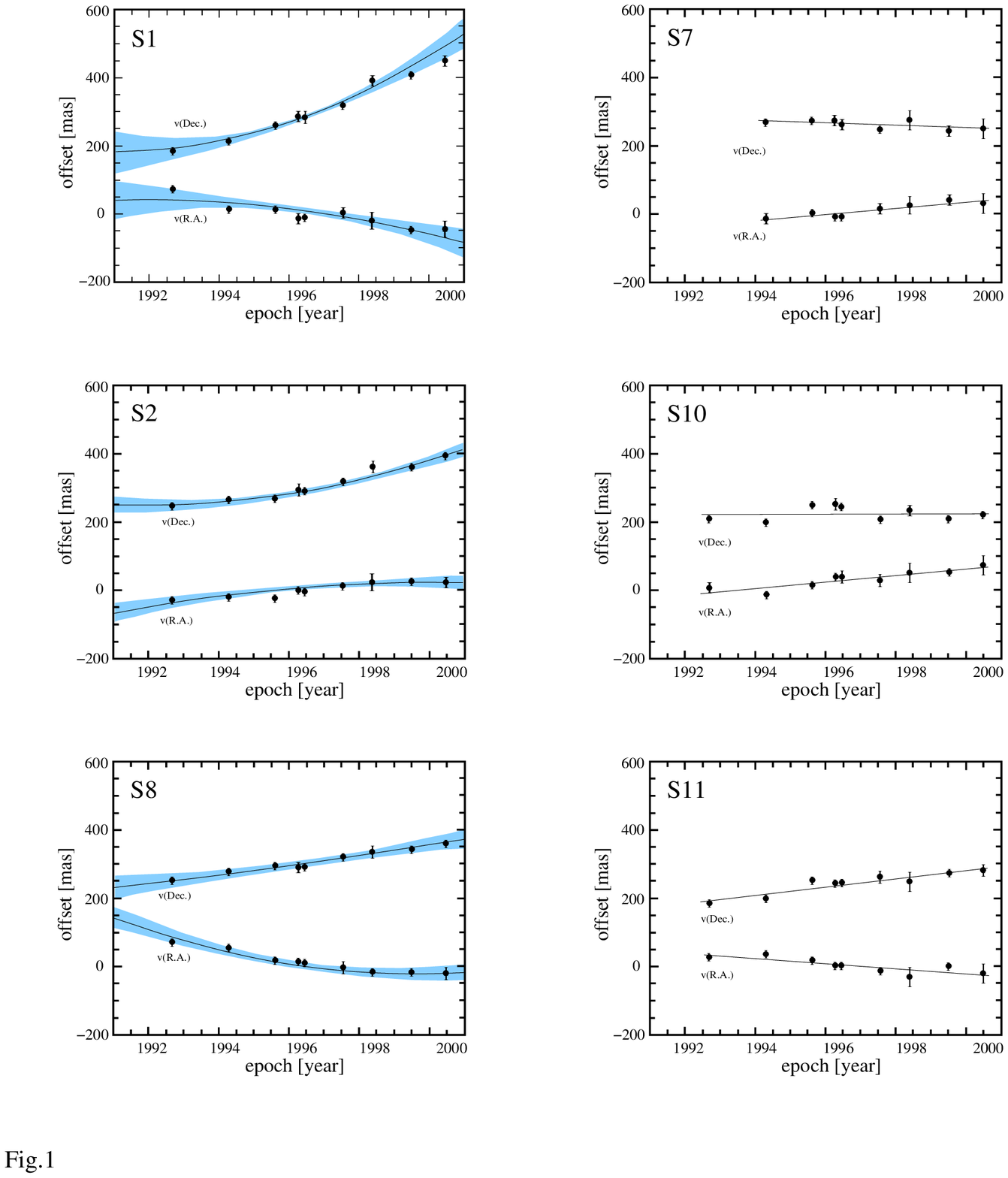,height=26cm,angle=-0} 
\end{tabular}
\end{center}
\end{figure} 
\newpage
\begin{figure}[!htp] 
\begin{center}
\begin{tabular}{c}
\psfig{figure=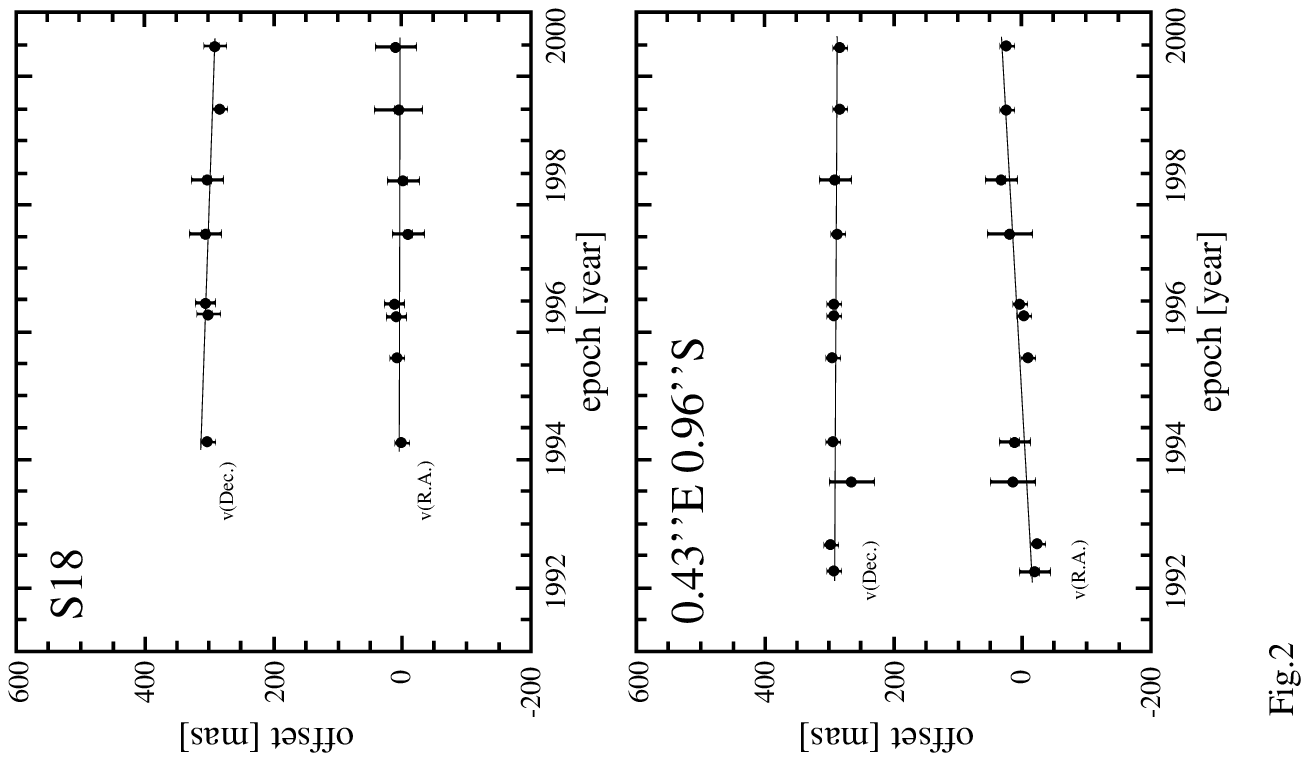,height=13cm,angle=-90} 
\end{tabular}
\end{center}
\end{figure} 
\newpage
\begin{figure}[!htp] 
\begin{center}
\begin{tabular}{c}
\psfig{figure=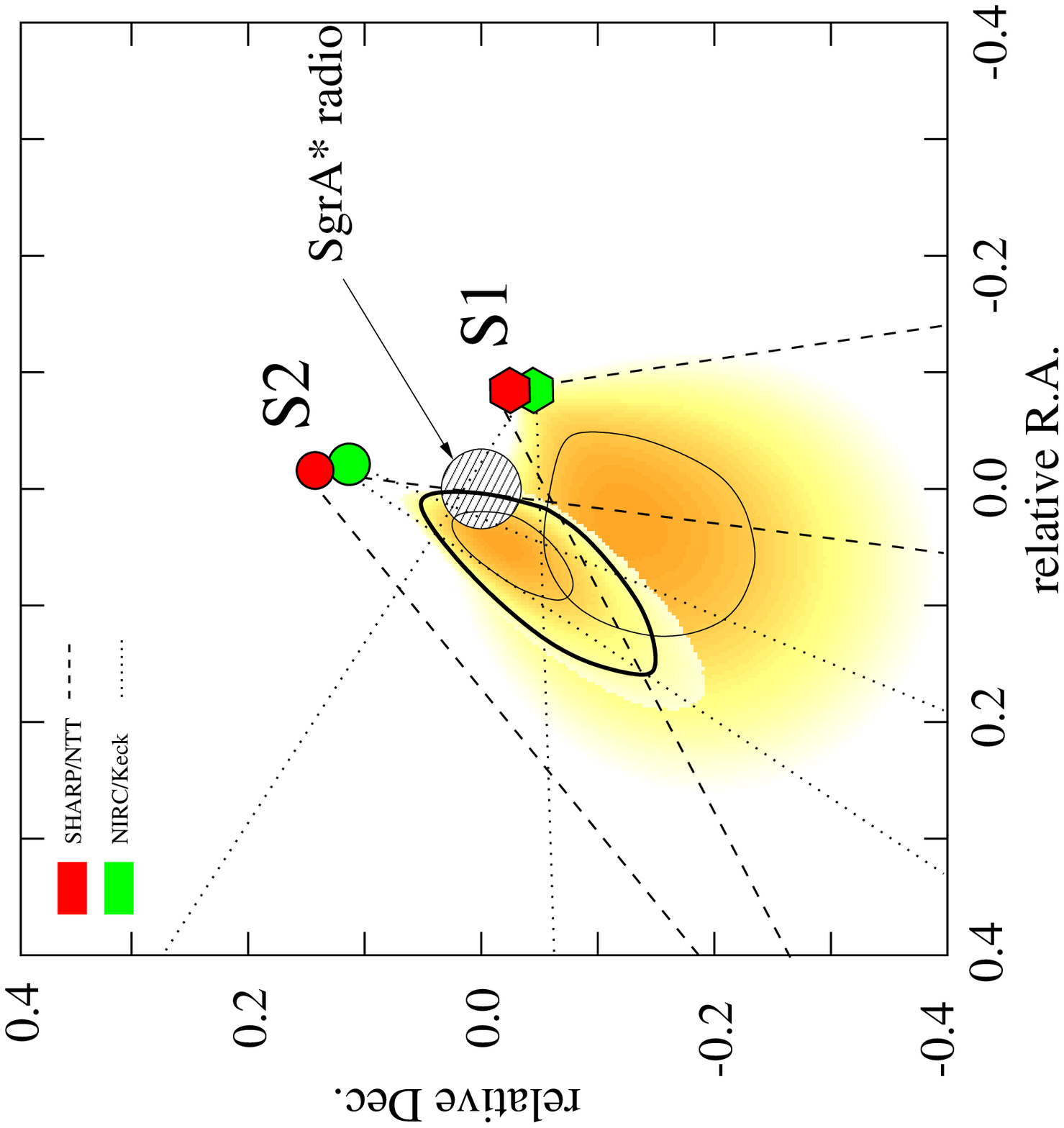,height=15cm,angle=-90} 
\end{tabular}
\end{center}
\end{figure} 
\newpage
\begin{figure}[!htp] 
\begin{center}
\begin{tabular}{c}
\psfig{figure=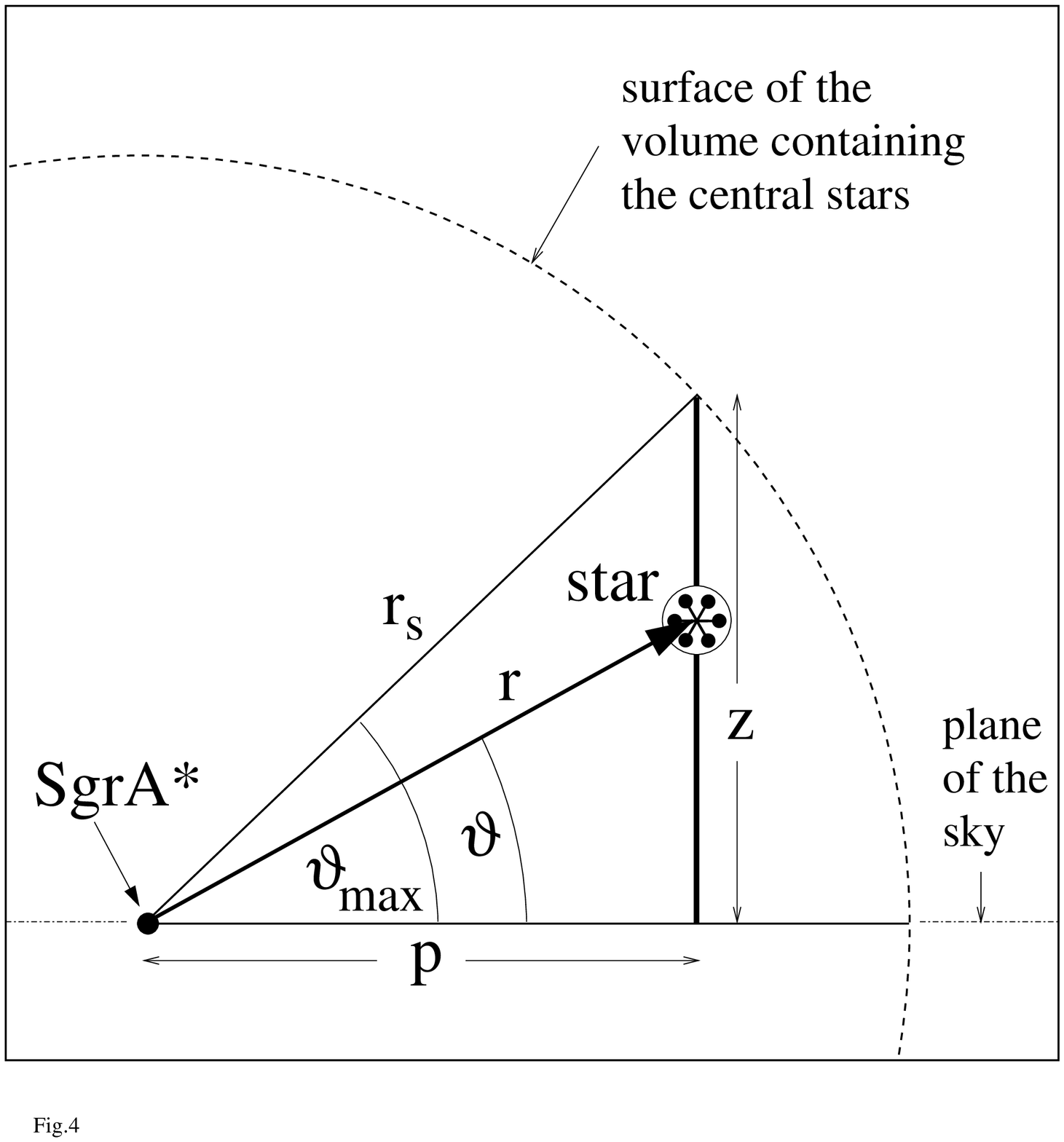,height=12cm,angle=-90} 
\end{tabular}
\end{center}
\end{figure} 
\newpage
\begin{figure}[!htp] 
\begin{center}
\begin{tabular}{c}
\psfig{figure=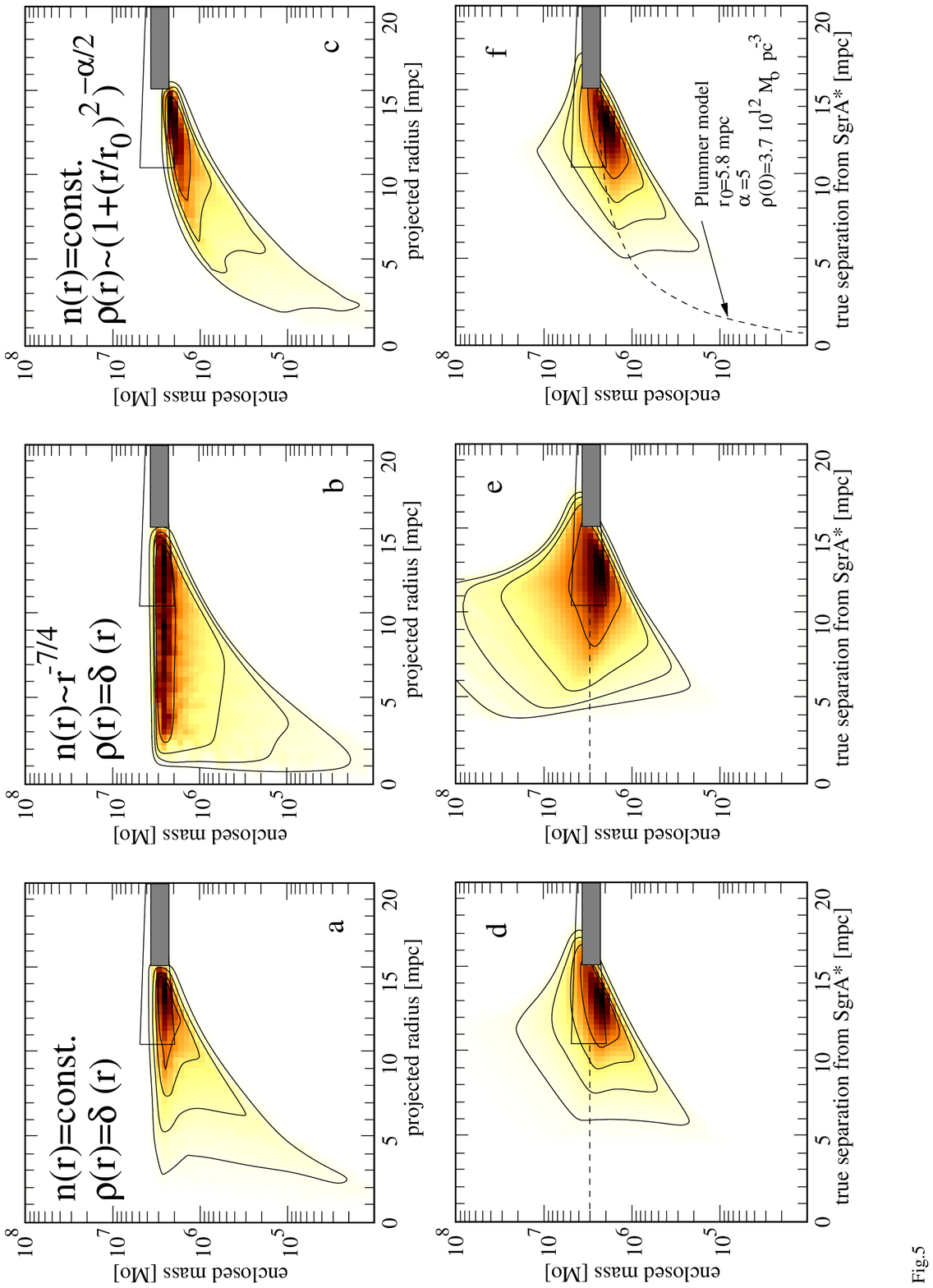,height=22cm,angle=0} 
\end{tabular}
\end{center}
\end{figure} 
\newpage
\begin{figure}[!htp] 
\begin{center}
\begin{tabular}{c}
\psfig{figure=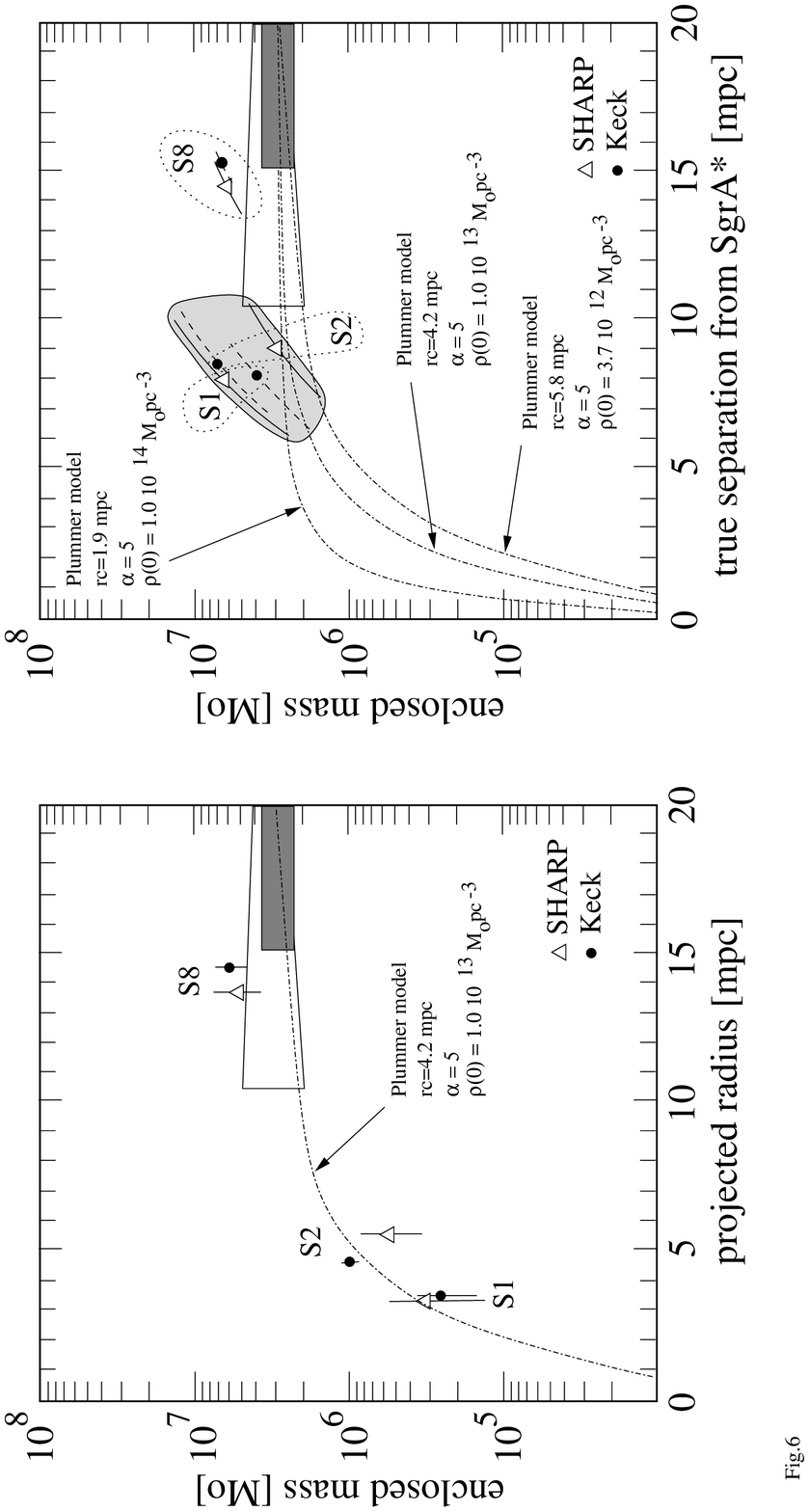,height=20cm} 
\end{tabular}
\end{center}
\end{figure} 
\newpage
\begin{figure}[!htp] 
\begin{center}
\begin{tabular}{c}
\psfig{figure=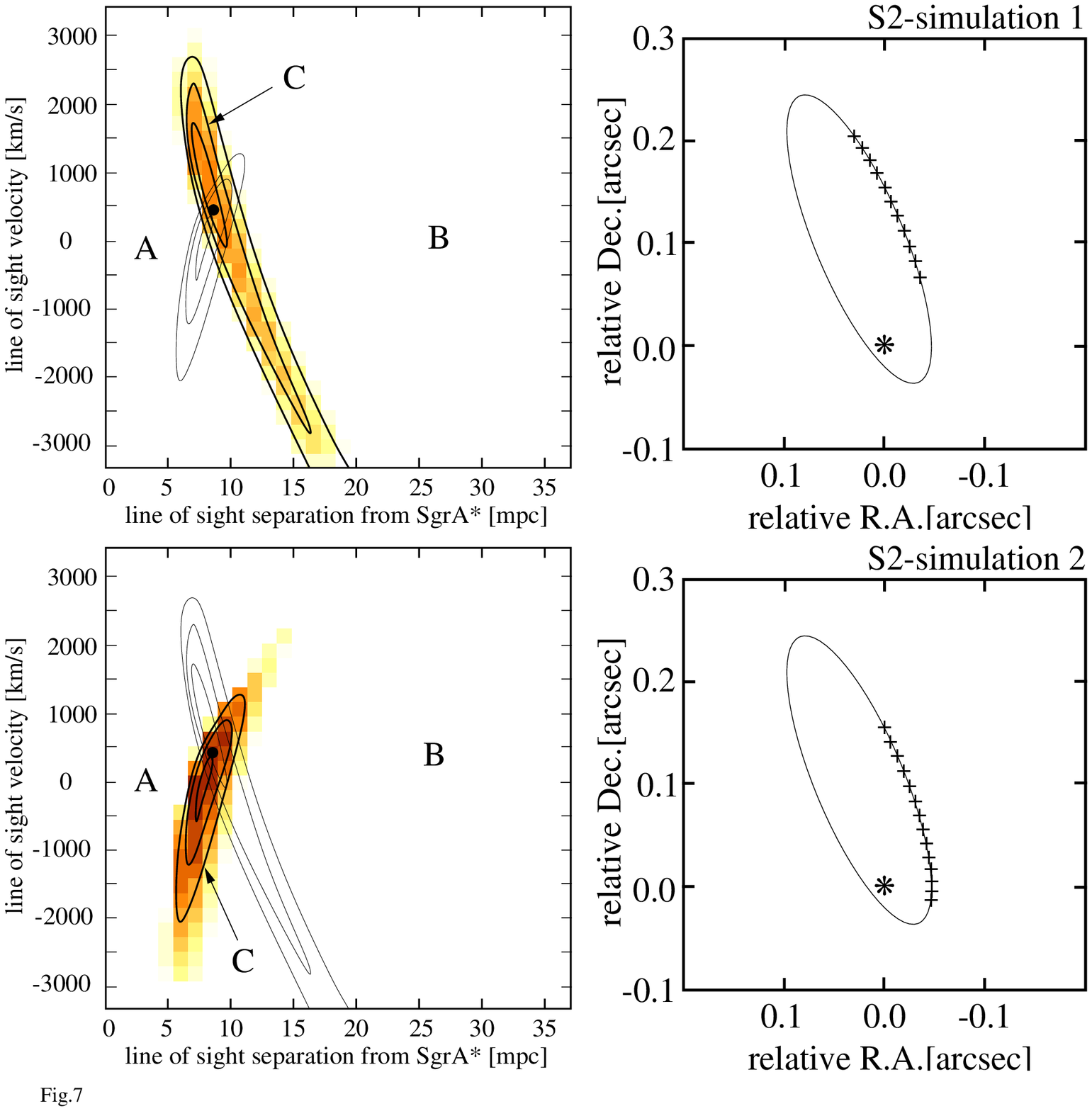,height=13cm,angle=-90} 
\end{tabular}
\end{center}
\end{figure} 
\newpage
\begin{figure}[!htp] 
\begin{center}
\begin{tabular}{c}
\psfig{figure=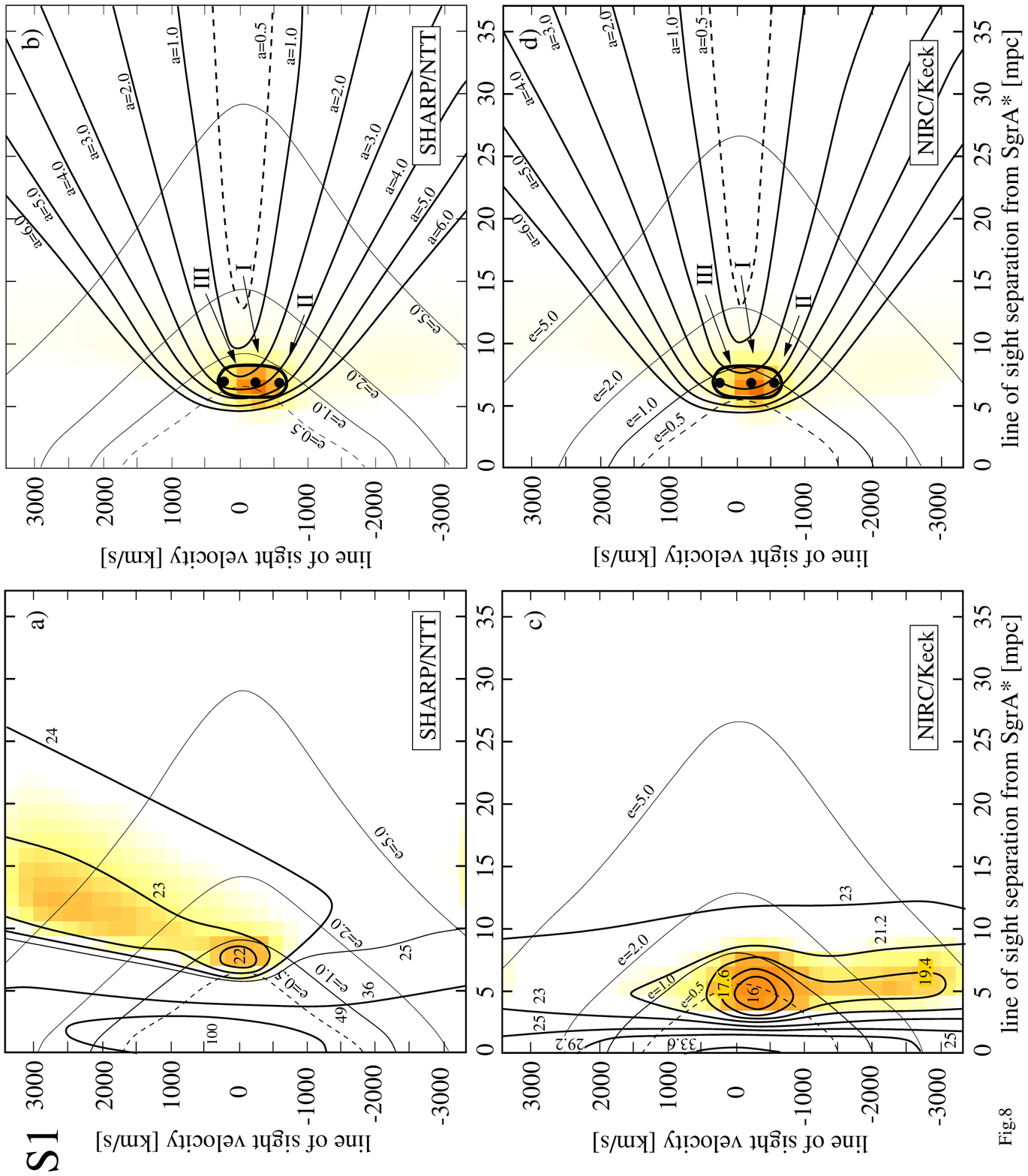,height=17cm,angle=-0} 
\end{tabular}
\end{center}
\end{figure} 
\newpage
\begin{figure}[!htp] 
\begin{center}
\begin{tabular}{c}
\psfig{figure=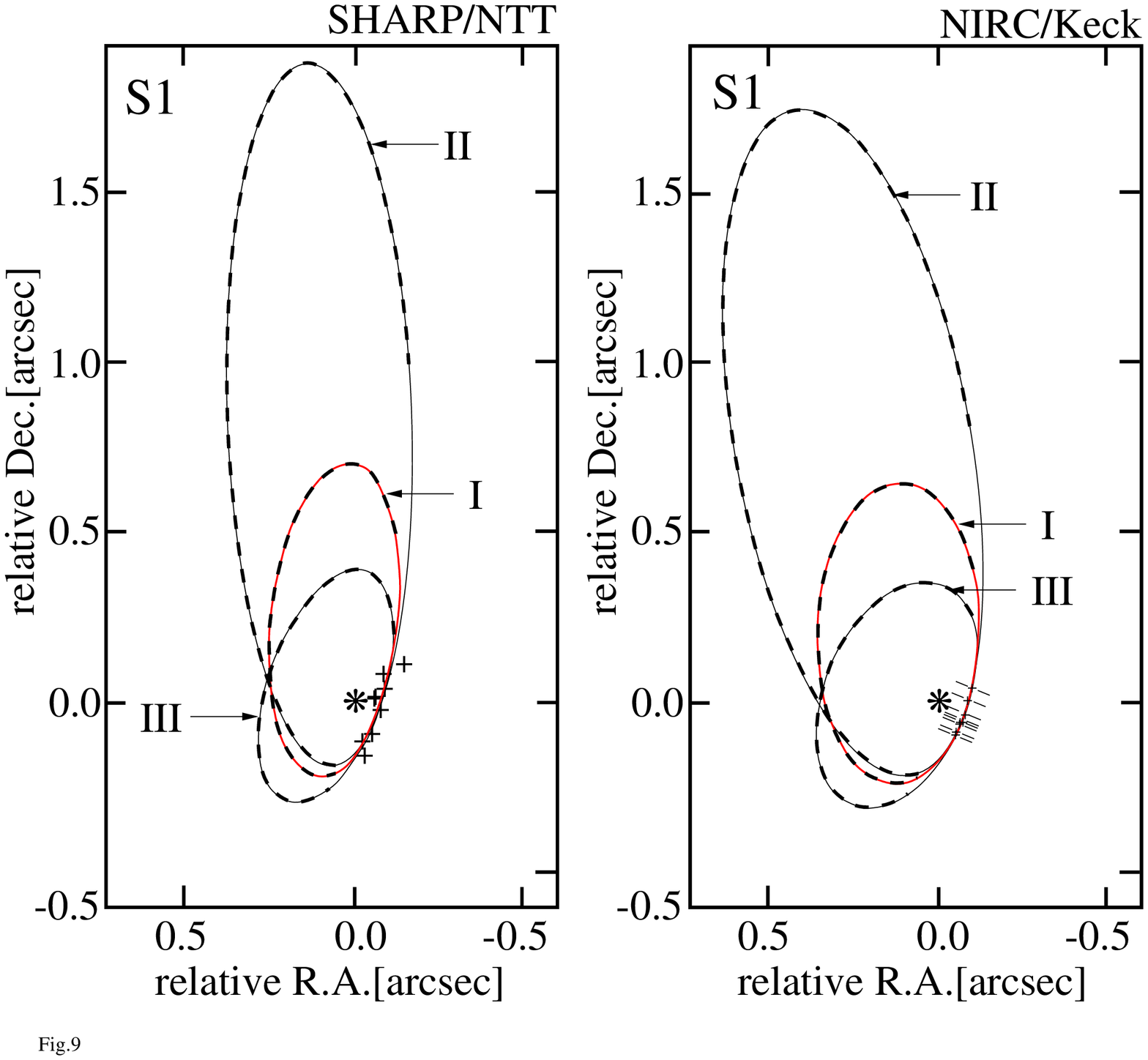,height=13cm,angle=-90} 
\end{tabular}
\end{center}
\end{figure} 
\newpage
\begin{figure}[!htp] 
\begin{center}
\begin{tabular}{c}
\psfig{figure=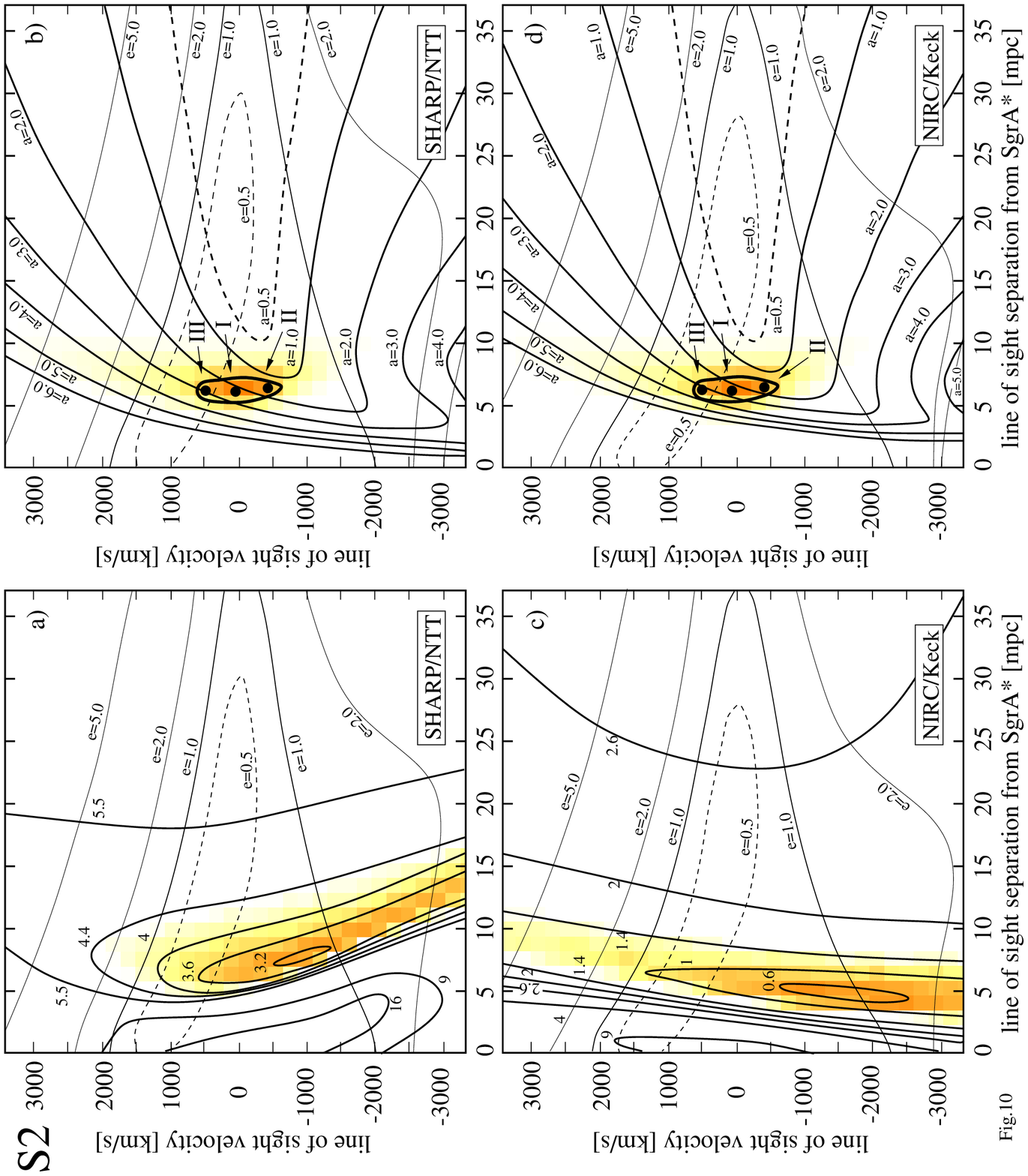,height=17cm,angle=-0} 
\end{tabular}
\end{center}
\end{figure} 
\newpage
\begin{figure}[!htp] 
\begin{center}
\begin{tabular}{c}
\psfig{figure=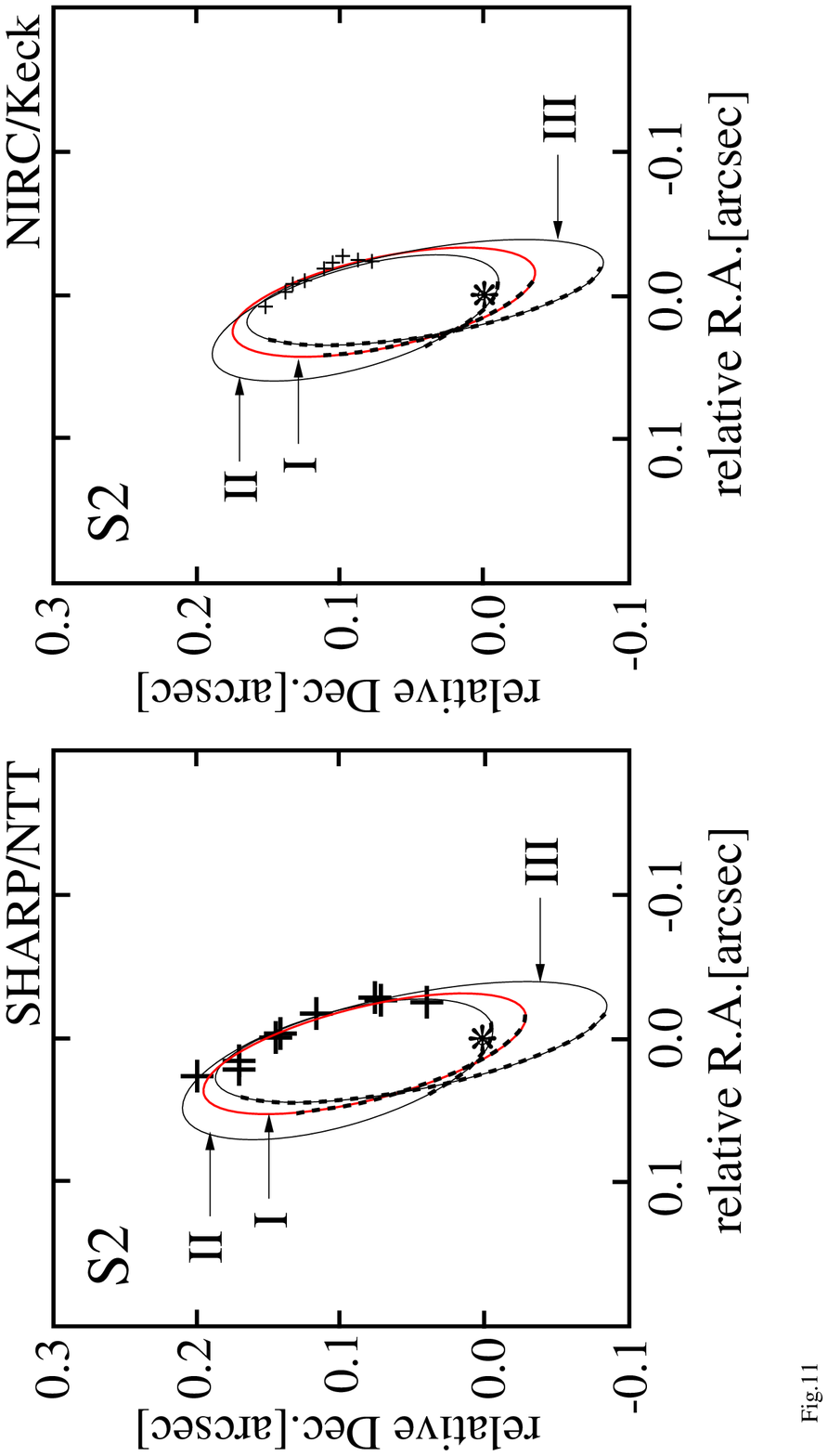,height=19cm,angle=-0} 
\end{tabular}
\end{center}
\end{figure} 
\newpage
\begin{figure}[!htp] 
\begin{center}
\begin{tabular}{c}
\psfig{figure=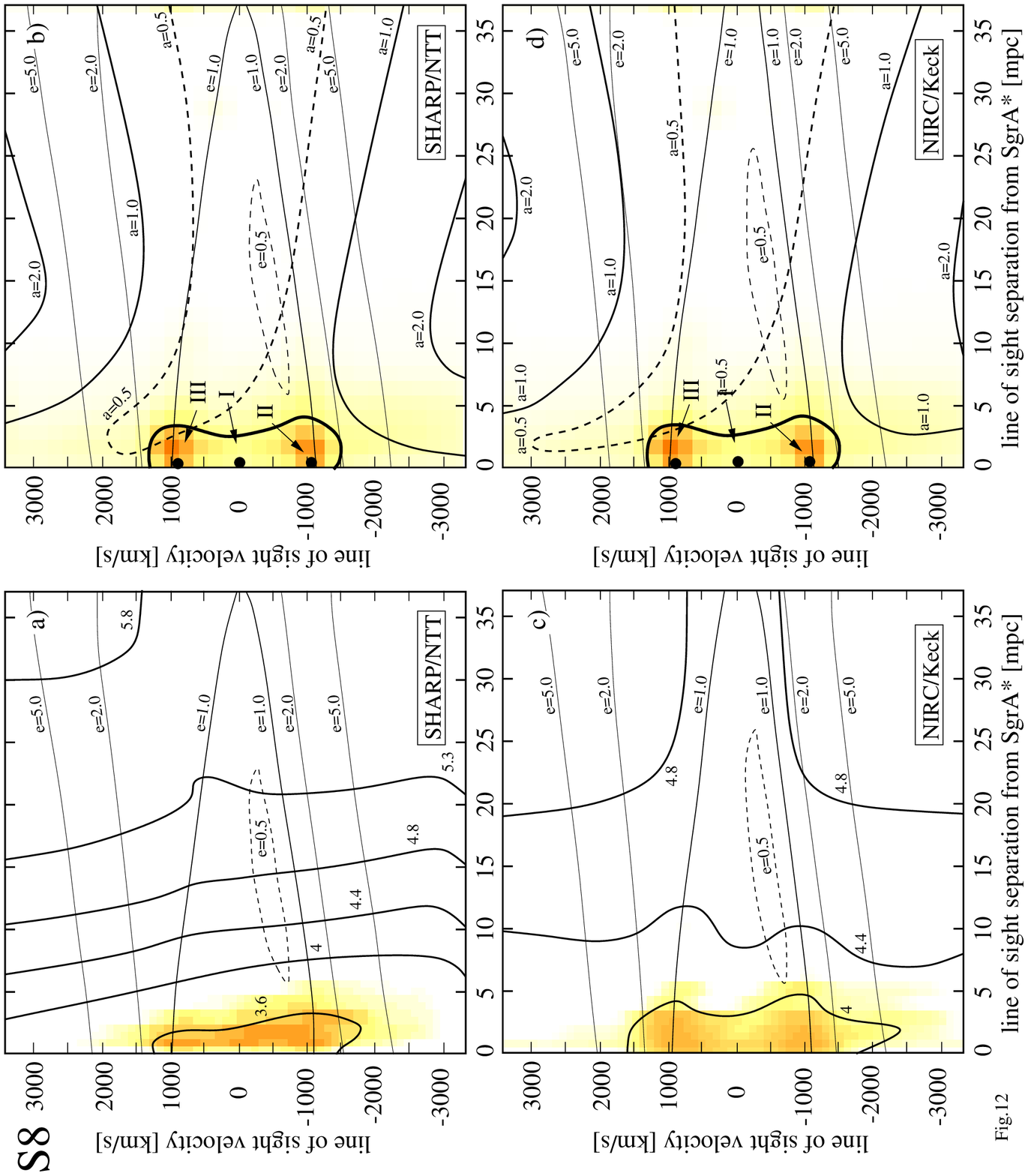,height=17cm,angle=-0} 
\end{tabular}
\end{center}
\end{figure} 
\newpage
\begin{figure}[!htp] 
\begin{center}
\begin{tabular}{c}
\psfig{figure=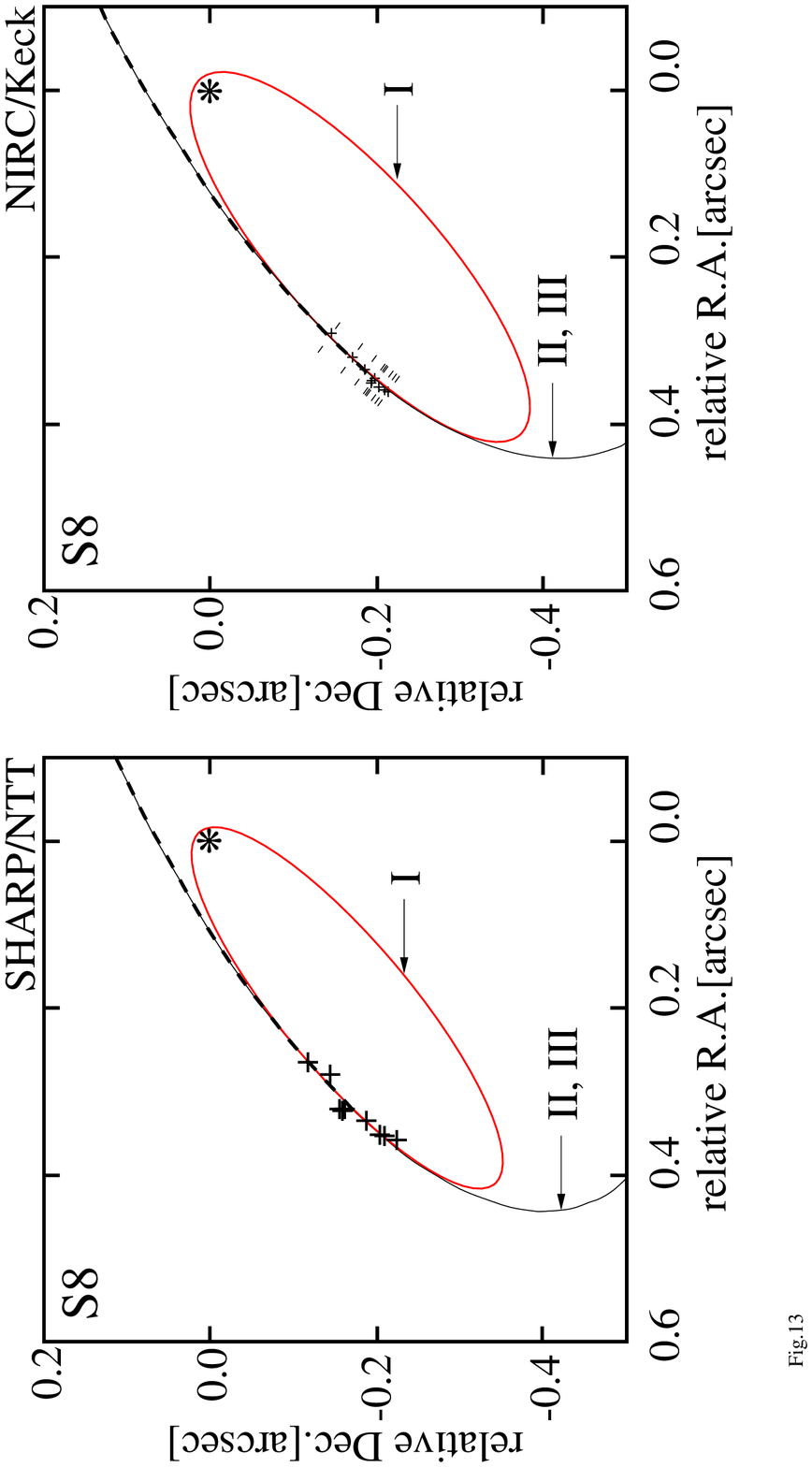,height=19cm,angle=-0} 
\end{tabular}
\end{center}
\end{figure} 
\newpage
\begin{figure}[!htp] 
\begin{center}
\begin{tabular}{c}
\psfig{figure=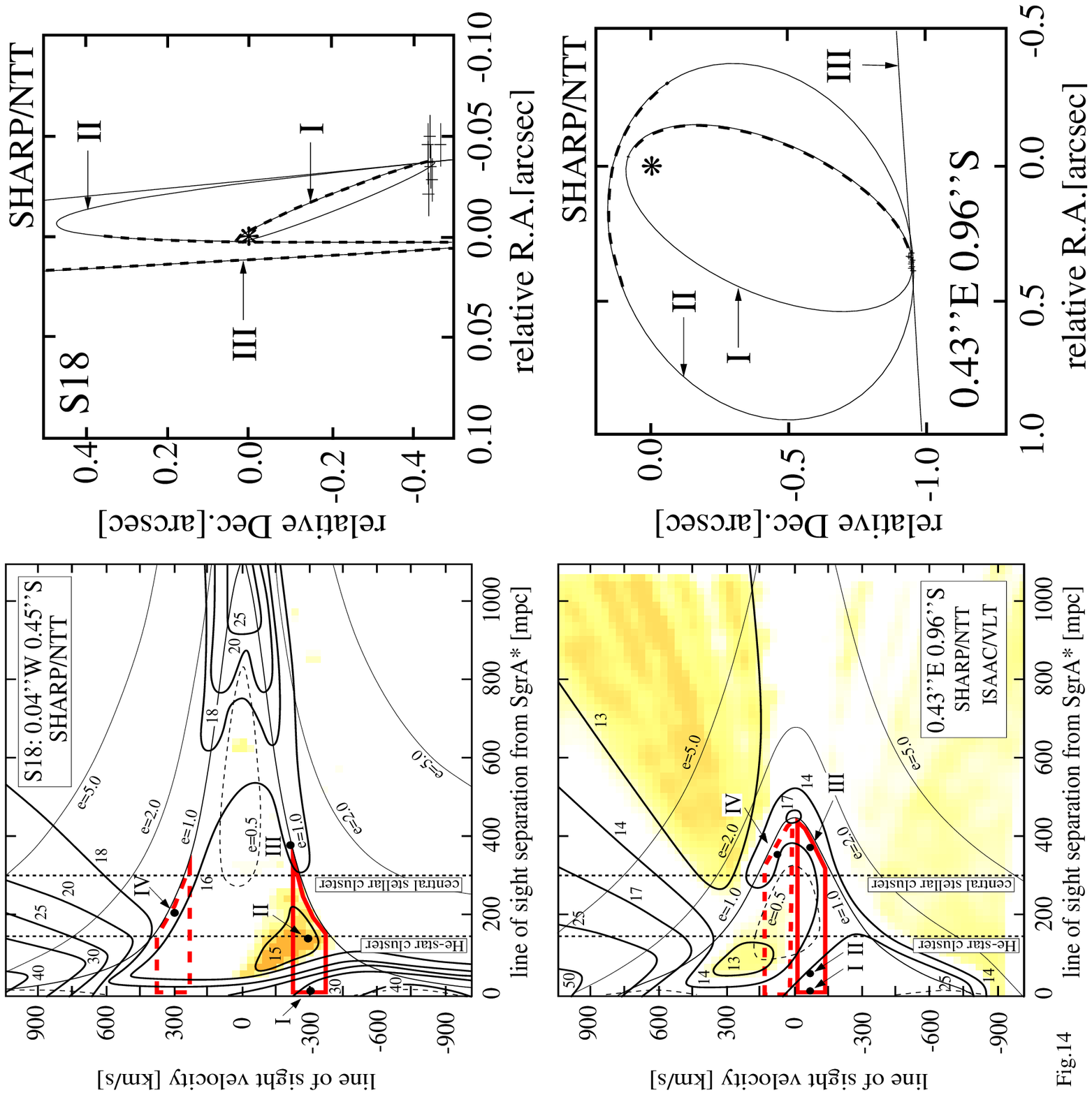,height=17cm,angle=-0} 
\end{tabular}
\end{center}
\end{figure} 
\newpage
\begin{figure}[!htp] 
\begin{center}
\begin{tabular}{c}
\psfig{figure=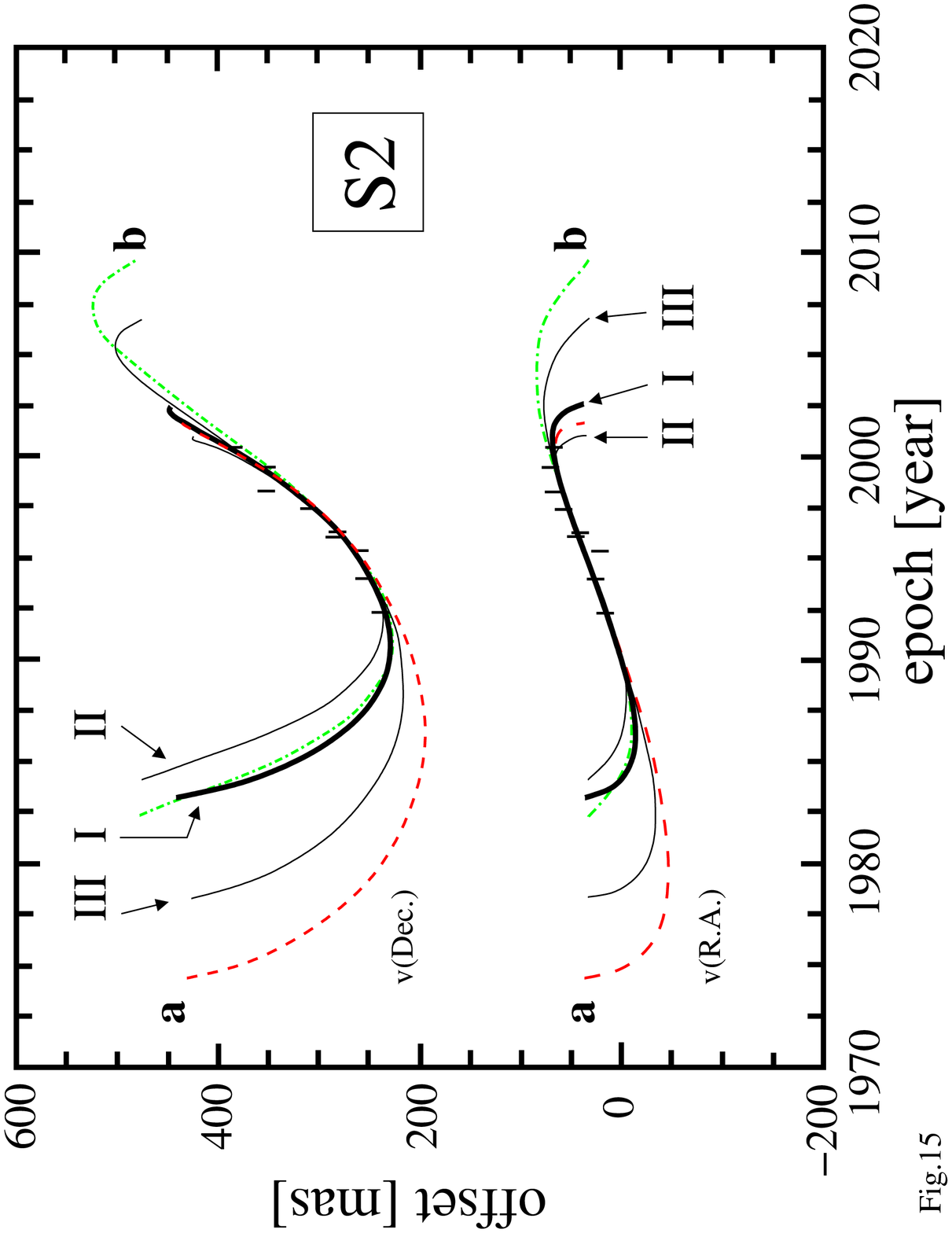,height=27cm,angle=-0} 
\end{tabular}
\end{center}
\end{figure} 
 
\clearpage
\appendix
\large
\begin{center}
{\bf APPENDIX}
\end{center}
\normalsize
Here we present the measured proper motion velocities and their transformation
to the epochs used for the orbit calculations. 
 
\begin{table}[!htb]
\caption{
\label{tapp01}
Comparison of Proper Motion Velocities }
\begin{center}
\begin{tabular}{rrrrrrrrr}\hline \hline
& SHARP & & Keck \\ 
& 1996.5 & & 1995.4 \\ 
& v$_{\alpha}$ & v$_{\delta}$ & v$_{\alpha}$ & v$_{\delta}$ \\ 
\\ \hline 
measured: S1 & +550$\pm$90 &-1440$\pm$100& +470$\pm$130& -1330$\pm$140 \\
S2 & -330$\pm$50 & -800$\pm$90 & -290$\pm$110& -500$\pm$50 \\
S8 & +490$\pm$60 & -560$\pm$40 & +720$\pm$80 & -530$\pm$110 \\
\hline \hline
transformed: S1 & +531 &-1480 & +489& -1291 \\
S2 & -249 & -587 & -371& -713 \\
S8 & +581 & -544 & +629& -545 \\
\hline \hline
mean: S1 & +541$\pm$60 &-1460$\pm$60 & +480$\pm$60& -1311$\pm$60 \\
S2       & -290$\pm$60 & -694$\pm$60 & -330$\pm$60&  -607$\pm$60 \\
S8       & +535$\pm$60 & -552$\pm$60 & +675$\pm$60&  -538$\pm$60 \\
\hline \hline
\end{tabular}
\end{center}
The velocities (in km/s) were obtained for 
epochs 1996.5 (SHARP) and 1995.4 (Keck; Ghez et al. 1998). 
We used the combined 1996.5 data in Tab.~\ref{t03}.
We combined the two data sets by transforming the velocities  
between epochs using the known accelerations.
For the combined (mean) proper motion velocities 
the mean (quadratic) difference between the R.A. and Dec. velocities 
of those two epochs are 80~km/s (95~km/s). 
As expected - this error accounts for velocity differences 
expected over 1.2 years from the measured curvatures of the order 
of about 80-160~km/s/yr as a mean per coordinate.
As an error for the mean velocities we used 60~km/s - obtained from
the mean quadratic deviation between the individually measured 
and transformed velocities for each of the 3 epochs in 
Tab.~\ref{tapp01} and Tab.~\ref{tapp02}.
\end{table}
 
\begin{table}[!htb]
\caption{
\label{tapp02}
Proper Motion Velocities for the 1997.6 Epoch}
\begin{center}
\begin{tabular}{rrrrrrrrr}\hline \hline
& SHARP to & & Keck to & & mean at \\ 
& epoch 97.6 & &epoch 97.6& & epoch 97.6\\ 
& v$_{\alpha}$ & v$_{\delta}$ & v$_{\alpha}$ & v$_{\delta}$ 
& v$_{\alpha}$ & v$_{\delta}$ \\ 
\\ \hline 
S1 & +606 &-1576 & +531& -1480 & +568$\pm$60 & -1528$\pm$60 \\
S2 & -293 & -879 & -249& -587  & -271$\pm$60 &  -733$\pm$60\\
S8 & +364 & -573 & +580& -544  & +472$\pm$60 &  -558$\pm$60 \\
\hline \hline
\end{tabular}
\end{center}
Transforming the velocities (in $km/s$) observed for 
epochs 1996.5 (SHARP) and 1995.4 (Keck)
to the time averaged Keck epoch 1997.6 (Ghez et al. 2000) using the 
projected accelerations in R.A. and Dec. as measured by SHARP (Tab.~\ref{t01}).
We used the combined 1997.6 data in Tab.~\ref{t03}.
For errors see caption to Tab.~\ref{tapp01}.
\end{table}

\end{document}